\documentclass[journal]{IEEEtran}
\ifCLASSINFOpdf
   \usepackage[pdftex]{graphicx}
\else
\fi
\usepackage[cmex10]{amsmath}
\usepackage[tight,footnotesize]{subfigure}
\usepackage{amsfonts}
\usepackage{graphics} 
\usepackage{setspace}
\usepackage{amssymb}
\usepackage{epsfig} 
\usepackage{stfloats}
\usepackage{multirow}
\usepackage{makecell}

\usepackage{algorithm}
\usepackage{algorithmicx}
\usepackage{algpseudocode}

\usepackage{amsthm}
\floatname{algorithm}{Algorithm}
\usepackage{enumerate}
\usepackage{subeqnarray}
\theoremstyle{plain}
\newtheorem{theorem}{Theorem}
\newtheorem{lemma}{Lemma}

\usepackage{bm}

\usepackage{color}
\newtheorem{remark}{Remark}
\begin{document}
%
\title{Gradient and Channel Aware Dynamic Scheduling for Over-the-Air Computation in Federated Edge Learning Systems}
\author{Jun~Du,~\IEEEmembership{Senior Member,~IEEE}, Bingqing Jiang,~\IEEEmembership{Student Member,~IEEE}, Chunxiao~Jiang,~\IEEEmembership{Senior Member,~IEEE}, Yuanming~Shi,~\IEEEmembership{Senior Member,~IEEE} and Zhu~Han,~\IEEEmembership{Fellow,~IEEE}
\thanks{This research was partially supported by the National Natural Science Foundation of China under Grants 61971257, partially by the Young Elite Scientist Sponsorship Program by CAST under Grant 2020QNRC001, and partially by NSF CNS-2107216, CNS-2128368, CMMI-2222810, Toyota and Amazon.
(\emph{Corresponding authors: Jun Du})}
\thanks{J. Du and B. Jiang are with the Department of Electronic Engineering, Tsinghua University, Beijing 100084, P. R. China (e-mail: jundu@tsinghua.edu.cn, jiangbq22@mails.tsinghua.edu.cn).}
\thanks{C. Jiang is with Tsinghua Space Center, Tsinghua University, Beijing 100084, P. R. China (e-mail: jchx@tsinghua.edu.cn).}
\thanks{Y. Shi is with the School of Information Science and Technology, ShanghaiTech University, China (e-mail: shiym@shanghaitech.edu.cn).}
\thanks{Z. Han is with the Department of Electrical and Computer Engineering at the University of Houston, Houston, TX 77004 USA, and also with the Department of Computer Science and Engineering, Kyung Hee University, Seoul, South Korea, 446-701. (e-mail: hanzhu22@gmail.com).}
}
\maketitle

\begin{abstract}
To satisfy the expected plethora of computation-heavy applications, federated edge learning (FEEL) is a new paradigm featuring distributed learning to carry the capacities of low-latency and privacy-preserving.
To further improve the efficiency of wireless data aggregation and model learning, over-the-air computation (AirComp) is emerging as a promising solution by using the superposition characteristics of wireless channels.
However, the fading and noise of wireless channels can cause aggregate distortions in AirComp enabled federated learning. In addition, the quality of collected data and energy consumption of edge devices may also impact the accuracy and efficiency of model aggregation as well as convergence.
To solve these problems, this work proposes a dynamic device scheduling mechanism, which can select qualified edge devices to transmit their local models with a proper power control policy so as to participate the model training at the server in federated learning via AirComp.
In this mechanism, the data importance is measured by the gradient of local model parameter, channel condition and energy consumption of the device jointly.
In particular, to fully use distributed datasets and accelerate the convergence rate of federated learning, the local updates of unselected devices are also retained and accumulated for future potential transmission, instead of being discarded directly.
Furthermore, the Lyapunov drift-plus-penalty optimization problem is formulated for searching the optimal device selection strategy.
Simulation results validate that the proposed scheduling mechanism can achieve higher test accuracy and faster convergence rate, and is robust against different channel conditions.
\end{abstract}


%
\IEEEpeerreviewmaketitle

\section{Introduction}

Future intelligent Internet-of-Things (IoT) networks are envisioned to experience a profound reformation from connecting massive things to connecting ubiquitous intelligence, which promotes the era of big data and puts forward stringent requirements of ultra-reliable and low latency communication (URLLC), energy efficiency, intelligence, and security.
In response, machine learning (ML) and artificial intelligence (AI) have been expected as important technologies to be widely introduced in wireless IoT networks~\cite{9606720, 9737357, du2020learningVTM}.
Due to their unique characteristics of universality, autonomy and adaptive optimization, ML and AI technologies can assist to support increasing multi-media and AI-empowered applications, by providing advanced wireless communications and computing capacities.
However, the astounding growth of AI applications and data volume generated by network edges bring heavy computation and communication workloads~\cite{du2020auctionTMC}. Especially for the traditional centralized training architecture relying on central or cloud servers, training computationally intensive tasks and transmitting such a mass of data will impose large latency and heavy traffic burden~\cite{lr}.
In addition, uploading local data to the server directly makes privacy security and preservation face significant challenges~\cite{9678323, 9188006}.
Aiming at these problems, distributed machine learning has been proposed to enable intelligent edges to train learning models collaboratively without uploading their local data to the server~\cite{9513259, 9296963, 9072647, 9103200}.

Federated learning was proposed by Google to enable collaborative training without sharing local datasets of edge devices, orchestrated by a server~\cite{9426913, 9460016}.
A typical federated learning is realized through a separated communication-and-computation approach. 
Specifically, each device runs the stochastic gradient descent (SGD) algorithm locally using the current global model, broadcasted by the server. 
Then device updates are aggregated at the server for the global model updating.
In this implementation, however, uploading local models over wireless channels might be resource-demanding, although the size of local models is much smaller than that of raw data observed locally.
On the other hand, training the global model at the server is to obtain certain kinds of functions of local models such as their weighted average instead of the individual model updates.
However, the traditional model aggregation in federated learning requires accurate timely signal reception and decoding of local model parameters, which results in unnecessary communication and computing resource consumption.
As an alternative, over-the-air computation (AirComp) has recently emerged as a promising solution to achieve fast and communication-efficient wireless data aggregation without precise distributed reception, especially for IoT applications~\cite{2007oac, 9502547}.

AirComp, as a combination scheme of communication and computation, has been shown as a communication-efficient approach to compute and train the target characteristic by exploiting the superposition property of wireless channels~\cite{oac1,Aircomp2022GCjiang}.
Different from the transmissions via wireless channels in federated learning,
AirComp enables a ``one-shot'' data aggregation from concurrent transmissions to harness the interferences
and the server receives the superimposed signals to be reconstructed, which is also aligned with the concept of secure aggregation~\cite{secure_aggregation}.
By designing appropriate pre-processing and post-processing functions at the edge devices and the server, respectively, AirComp can be well-suited to the federated edge learning (FEEL) systems regarding to high communication efficiency and low latency demands. 
In addition,
efficient device scheduling mechanisms, including power control, device selection, resource allocation, etc., can be realized to improve the accuracy and efficiency of model aggregation in FEEL systems~\cite{transmission2022Cao, scheduling2, scheduling3}.
In this work, we will investigate the dynamic device control and management mechanisms aiming to optimize the synthetic performance, including training accuracy, energy efficiency, etc., of the designed AirComp enabled FEEL system.

\subsection{Contributions and Organization}

In this work, we will establish an AirComp enabled FEEL system, in which several distributed edge devices and an edge server are organized to operate a shared global training task collaboratively and communication-efficiently.
The main contributions of the paper are summarized as follows.

\begin{itemize}
  \item For a computation-efficient AirComp in FEEL systems, we propose a gradient and channel aware device scheduling mechanism to select devices, which have satisfied data qualities, channel status, and energy consumptions based on a delicately defined \emph{indicator of device quality}. In addition, the power control mechanism based on the channel inversion is designed to resist channel fading. Then a Lyapunov-drift-minimization-based optimization algorithm is designed to solve the device scheduling problem through a distributed and online approach.	
  \item We design an accumulated residual feedback based gradient transmission mechanism that allows the edge server to utilize local updates of unselected devices sufficiently. Compared with traditional strategies of discarding these updates directly, the proposed mechanism can improve the efficiency and accuracy of global model aggregation and convergence. Moreover, the convergence bound of this mechanism in the FEEL system is also characterized.
  \item We analyze the performance of the proposed device scheduling mechanism and run simulations on MNIST and CIFAR-10 datasets. Simulation results validate that the proposed mechanism achieves significant improvements on test accuracy, energy consumption, as well as the adaptation and efficiency in non-independently identical distributed (non-i.i.d.) dataset scenarios, compared with baselines introduced from state-of-the-art studies.
\end{itemize}

The rest of this paper is organized as follows.
Related works are reviewed in the following subsection.
Section~\ref{smlm} sets up the system model and learning framework.
In Section~\ref{residual}, a dynamic residual mechanism is designed and its convergence analysis is provided.
A gradient and channel aware dynamic device scheduling mechanism for AirComp enabled FEEL is proposed in Section~\ref{mechanism}.
Simulation results are shown in Section~\ref{simulation}, and finally conclusions are drawn in Section~\ref{conculsion}.

\subsection{Related Work}

Federated learning is becoming an emerging paradigm for enabling ML and AI at the wireless edge devices, in which edge devices train a global model collaboratively, and upload their local models by applying orthogonal multiple access based resource allocation schemes, such as time division multiple access (TDMA) and orthogonal frequency division multiple access (OFDMA).
However, the required radio resources are linearly scaling with a large number of edge devices with
limited bandwidth and power, which introduces a substantial communication latency and resource consumption~\cite{distributed_learning}.
Therefore, it is important to investigate communication-efficient mechanisms for the realization and implementation of a federated learning framework with high training accuracy, energy efficiency and low latency.

Recently, there have been significant efforts to incorporate physical and network layer characteristics of wireless networks into communication-efficient federated learning.
In order to optimize the training speed and model accuracy, an accelerated gradient-descent multiple access algorithm was proposed in~\cite{algorithm1} by using momentum-based gradient signals over multiple access noisy fading multiple access channels.
Under the resource constraint, the authors of~\cite{localgloabal} proposed a learning control algorithm with the best tradeoff between local updates and global parameter aggregation.
Considering the redundant observations and diversity of devices in terms of local datasets, channel conditions, energy status, etc., many studies have paid attention to designing efficient device scheduling policies, according to which only parts of devices are selected and scheduled to participate the global model updating in each iteration.
In~\cite{fl-privacy}, the authors
designed a privacy-preserving and communication-efficient device scheduling mechanism, in which both differential privacy and communication costs were considered to make less frequent communications with fewer devices participation.
The authors of~\cite{allocation1} optimized the user selection and resource allocation mechanism jointly to minimize the training loss of federated learning.
In~\cite{divFL}, the authors proposed a communication-efficient device selection strategy by selecting a set of more informative devices, which can improve the convergence speed, fairness and learning efficiency.
A framework for low-latency broadband federated learning was investigated in~\cite{scheduling7} by taking channel state information into consideration.
In addition, the channel conditions and local model updates were both formulated in~\cite{scheduling5} to develop a device scheduling and resource allocation mechanism in federated learning.
However, most current studies only took the channel condition and model update into consideration, ignoring the energy efficiency and lacking a comprehensive evaluation and modeling of device quality.
Moreover, in the studies above, the model update information of unselected devices was discarded, which led to an incomplete utilization of data samples observed in FEEL systems.

The research of AirComp mainly focuses on addressing the channel distortion. 
Some novel works have considered designing the transmitter and receiver policy with the objective of minimizing the computation mean-square error (MSE), based on which a sum-power constraint was considered to better distribute the power to the devices in~\cite{sum_energy}.
Also, the authors of~\cite{broadband} proposed a novel M-frequency AirComp system to accommodate a large scale of devices,
and spatial-and-temporal correlated signals were considered in~\cite{correlated} for a more practical and realistic scenario.
In recent years, the AirComp technique has been integrated into federated learning to improve the learning performance referring to radio resource and energy efficiency, training convergence and accuracy, etc.
In~\cite{power1}, a power control policy for AirComp enabled federated learning was investigated by introducing the gradient statistics.
A modified federated learning algorithm was proposed in~\cite{lr} to optimize the local learning rates and adapt the fading channel.
Based on the convergence behavior in AirComp enabled federated learning, the authors of~\cite{transmission2022Cao} proposed a novel power control policy to combat against the aggregation errors, which enhanced the training accuracy and accelerated the training speed greatly.
In~\cite{scheduling1}, an energy-aware dynamic device scheduling algorithm was designed to optimize the training performance, considering both communication and computation energies of local training.
In summary, AirComp, as a nascent technology revolutionizing the traditional communication-and-computation separation method, has provided a desirable possibility of combining communication and computation together to realize a communication-efficient distributed intelligence.


\section{System Model and Learning Mechanism}
\label{smlm}

\begin{figure}[!t]
  \centering
  \includegraphics[width=0.479\textwidth]{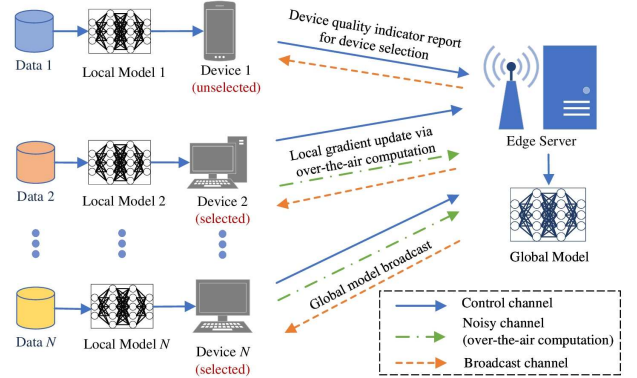}
  \caption{Architecture of AirComp enabled FEEL system.}\vspace{-3mm}
  \label{fig:system}
\end{figure}

This work considers a FEEL system consisting of an edge server and a finite set of $N$ decentralized edge devices equipped with single-antennas, denoted by $  \mathcal{N} \!=\! \left\lbrace  1,2, \!\cdots\! , N \right\rbrace  $, as illustrated in Fig.~\ref{fig:system}.
Each device $n\in \mathsf{\mathcal{N}}$ collects labeled dataset $\mathsf{\mathcal{D}}_n$ locally and then trains a shared global model collaboratively, such as a classifier or identifier, by interacting with the edge server.
Note here that the local models refer to the trained models at edge devices, and the global model refers to the model aggregated at the edge server.
Before proceeding further, we summarize the main notations used throughout the following sections in Table~\ref{table:notations} for
convenience.

Next, we establish the gradient updating oriented federated learning model in Section~\ref{fed_model}. To realize the gradient aggregation through a communication-efficient approach, an AirComp-based transmission and fusion policies are designed in Section~\ref{aircomp}. Then the detailed implementation of the AirComp enabled FEEL mechanism is proposed in Section~\ref{model_learning}.

\subsection{Federated Learning Model}
\label{fed_model}

Instead of uploading private and original data to the edge server, device $n$ in the FEEL system as shown in Fig.~\ref{fig:system} only has to upload the model parameters or its gradients obtained by its local training.
Then a model aggregation and updating will be executed at the server.
In this work, \textit{the gradient of training model parameters} is expected to be uploaded and aggregated by the edge devices and the edge server, respectively, to learn the global model.
Comparing with traditional centralized data aggregation and training, such a federated learning mechanism can reduce the communication overhead between edge devices and the server, alleviate the computation workload at the server, and avoid privacy disclosure.

To facilitate the supervised federated learning, we adopt cross-entropy $ l(\mathbf{w}; \mathbf{s}_{i}, q_{i}) $ as the loss function of data sample $\left( \mathbf{s}_{i}, q_{i} \right) $, which measures the prediction error between model parameter $ \mathbf{w} $ on training sample $\mathbf{s}_{i}$ with respect to its ground-truth label $q_{i}$. Then the local loss function of device $ n $'s model parameter $ \mathbf{w} $ on $\mathsf{\mathcal{D}}_n$ ($\forall n\!\in\! \mathsf{\mathcal{N}}$) can be given by
\begin{equation}
f_n\left( \mathbf{w} \right) = \frac{1}{\left| \mathsf{\mathcal{D}}_n\right|  }\sum \nolimits_{\left( \mathbf{s}_{i}, q_{i} \right) \in \mathsf{\mathcal{D}}_n}
{f \left(\mathbf{w};\mathbf{s}_{i},q_{i} \right)},
\end{equation}
where $\left| \mathsf{\mathcal{D}}_n\right|$ denotes the size of dataset $\mathsf{\mathcal{D}}_n$.
Considering different edge devices collect dataset independently, we assume that $\mathsf{\mathcal{D}}_m \cap \mathsf{\mathcal{D}}_n =\emptyset$ if $m \neq n$.
Then the global loss function on all distributed dataset $\left\{\cup_{n \in \mathsf{\mathcal{N}}}\mathsf{\mathcal{D}}_n  \right\}$ can be calculated as
\begin{equation}
f\left( \mathbf{w} \right) =\frac{1}{\left|\cup_{n \in \mathsf{\mathcal{N}}}\mathsf{\mathcal{D}}_n\right|}\sum\limits_{n=1}^{N}\left| \mathsf{\mathcal{D}}_n\right|f_n\left( \mathbf{w}\right),
\label{equ:gl}
\end{equation}
where $\left|\cup_{n \in \mathsf{\mathcal{N}}}\mathsf{\mathcal{D}}_n\right|=\sum\nolimits_{n=1}^{N}\left| \mathsf{\mathcal{D}}_n\right|$.
The global loss function summarized in (\ref{equ:gl}) measures that how well the model fits the entire corpus of data on average. Then the objective of federated learning is to find the best model parameter $ \mathbf{w}^{*} $ which can minimize the global loss function, i.e.,
\begin{equation}
\label{objective}
\mathbf{w}^*= arg \min_{\mathbf{w}}f(\mathbf{w}).
\end{equation}

\begin{table}[!t]
	\renewcommand\arraystretch{1}
	\centering
	\footnotesize
	\caption{List of Main Notations in the FEEL System.}
	\begin{tabular}{c|l}
		\hline \hline
		\textbf{Parameter} & \textbf{~~~~~~~~~~~~~~~~~~~~~~~~~~~Definition} \\
		\hline
		$\mathbf{\mathcal{D}}_n$/$\mathbf{\mathcal{D}}^m_n$ $$ & local dataset/local mini-batch\\
		$l\left(\mathbf{w};\mathbf{s}_i,q_i\right) $ & loss function\\
		$\mathbf{w}_t$ & global model parameter in round $t$\\
		$ \left( \mathbf{s}_i,q_i\right)  $ & data sample\\
		$ f\!\left( \mathbf{w} \right)  $/$ f_n\!\left( \mathbf{w}\right)  $ & global loss function/local loss function of device $n$\\
		$ \mathbf{S}_t $ & subset of selected devices in round $t$\\
		$ a_{n,t} \!\in\! \left\lbrace0,1 \right\rbrace $ & selection stratigy on device $ n $ in round $ t $\\
		$ h_{n,t} $ & channel gain between device $ n $ and server in round $ t $\\
		$ \sigma_t $ & power scaling factor in round $ t $\\
		$ P_{n,t} $ & transmit power of device $n$ in round $ t $\\
		$ \mathbf{g}_{t} $/$ \mathbf{g}_{n,t} $ & global/local model gradient (of device $ n $) in round $ t $\\
		$ \mathbf{r}_{n,t} $ & accumulated residual of device $ n $ in round $ t $\\
		$ \tilde{\mathbf{g}}_{n,t} $ & combination of $ \mathbf{g}_{n,t} $ and $ \mathbf{r}_{n,t} $\\
		$ \gamma_s $/$ \gamma_\text{thr}$ & received SNR/SNR threshold at edge server\\
		$ \mathbf{\varepsilon}_{t} $ & additive white Gaussian noise vector ($ \sigma_0^2 $)\\
		$ \zeta_t $ & learning rate in round $ t $\\
		$ E_{n,t} $ & transmit energy of device $ n $ in round $ t $\\
		$ G^2  $ & variance of local model gradient\\
		$ Pr_t  $ & average probability of unselection in round $ t $\\
		$ v_{DSI/CSI_{n,t}}  $ & indicator of DSI/CSI of device $ n $ in round $t$\\
		$ V_{n,t} $ & device importance of device $ n $ in round $ t $\\
		$ \rho_1  $/$ \rho_2  $ & weight of channel condition/local update significance\\
		$ I_{n,t} $ & device quality of device $ n $ in round $ t $\\
		$ \lambda_E  $/$ \lambda_V  $ & weight of energy consumtion level/device importance\\
		$ \alpha $ & Lyapunov factor\\
		$ L_t $ & Lyapunov function\\
		$ \Delta_T $ & $ T $-round Lyapunov drift\\
		\hline\hline
	\end{tabular}
	\label{table:notations}%
\end{table}%

\subsection{Over-the-air Gradient Aggregation}
\label{aircomp}

The AirComp technique enables an efficient gradient fusion from concurrent transmissions of device updates.
In this work,
we establish an AirComp based gradient data collection mechanism to adopt for uplink channel access from each edge device to the server.
We consider
a single communication round of AirComp in a globe model updating cycle, indexed by $t$, $t\in \left\{ 1,2,\cdots T\right\}$, and $T$ is the total number of communication rounds for the global model training.
In this process, only a part of devices with appreciated channel conditions and data qualities will be selected to upload their gradient data for an efficient and reliable model aggregation at the server~\cite{du2017contractJSAC}.
Denote subset $\mathbf{S}_t$ as the selected devices in the $t_{th}$ communication round.
Then the quantity of these devices selected can be represented by  $ \left| \mathbf{S}_t\right| =\sum_{n=1}^{N}a_{n,t} $, where $ a_{n,t}\in\left\lbrace 0,1\right\rbrace  $ is a binary variable indicating that whether device $n$ is selected or not for gradient aggregation. Specifically, $ a_{n,t}= 1 $  denotes the case where device $ n $ is selected in round $ t$, and otherwise, $ a_{n,t} =0 $. Therefore the element of $\mathbf{S}_t$ can be defined as $ S_{n,t} \! =\! \left\lbrace a_{n,t}, n\in \mathcal{N} |a_{n,t}\! =\! 1 \right\rbrace $.
According to~(\ref{objective}), we aim to minimize the expectation of final training loss by optimizing device selection strategy $ a_{n,t} $, given the total number of training rounds $T$, i.e., 	
\begin{equation}
\label{obj}
\min_{\left\{a_{n,t}\right\}_{t=1}^T} E\left[ f_{\mathbf{w}_{T}}\right],
\end{equation}
After being selected, these edge devices will share one public radio resource block to transmit their gradients of local model parameters simultaneously to implement AirComp.
By employing the superposition property of multiple-access channels,
the edge server will receive a weighted sum of gradient signals, which refers to the channel gains combined with power control policies.
Such an AirComp mechanism above integrates communication and computation, so as to achieve a low-latency and low-complexity data aggregation in the FEEL system.
In this work, we assume that transmissions of gradient data are strictly synchronized among all the selected devices.
Let $ h_{n,t} \in \mathbb{C}$ be the channel gain between device $ n $ and the edge server in communication round $t$, which is considered as quasi-static within a radio resource block. To be specific, $ h_{n,t} $ is stable in a single block but varies over different blocks and communication rounds.
To resist the channel fading effectively, we design a power control mechanism based on the channel inversion policy proposed in~\cite{scheduling7, feel, fedsplit}. Denote $ \sigma_t $ as the power scaling factor, which determines the received signal-to-noise ratio (SNR) at the server.
Then transmit power $ P_{n,t} $ of device $n$ in round $t$ should be set as
\begin{equation}
\label{equ:p}
P_{n,t}=\sigma_t\frac{h_{n,t}^H}{\left| h_{n,t}\right| ^2},
\end{equation}
where $ h_{n\!,t}^H $ is the conjugate of $ h_{n\!,t} $.
Let $ \mathbf{g}_{n\!,t} \!=\! \left[g_{n,1},\!\cdots\! ,g_{n,L_g}\!\right]^T $ denote the local model gradient vector of device $ n $ in round $t$,
and $ L_g $ is the number of local gradient elements.
Based on the designed power control policy, transmission signal $ \mathbf{x}_{n,t} $ sent by device $n$ can be represented as
$ \mathbf{x}_{n,t}\!=\!\Phi\left( {\mathbf{g}}_{n,t}\right) $,
where $ \Phi\left( \cdot \right)  $ denotes a pre-processing and normalization function to ensure that $ \mathbf{x}_{n,t} $ has zero-mean and $P_{n,t}^2$-variance, i.e., $ E\left[ \mathbf{x}_{n,t}\mathbf{x}^H _{n,t}\right] \!=\!E\left[ \left\| \mathbf{x}_{n,t}\right\|_2^2\right]\!=\!P_{n,t}^2 $~\cite{scheduling7}.
Similarly, $ \Phi^{-1}\left( \cdot \right) $ is a designed post-processing and denormalization function performed at the edge server.
Then at the edge server, the received signal $\mathbf{y}_t$ from all selected devices via AirComp can be represented as
\begin{equation}
\mathbf{y}_t =\sum\nolimits_{n\in \mathbf{S}_t}h_{n,t}\mathbf{x}_{n,t}+\bm{\varepsilon}_t
\label{equ:y_t}
\end{equation}
where $ \bm{\varepsilon}_t \in \mathbb{R}^{L_g} $ is the additive white Gaussian channel noise vector with variance $\sigma_0^2$. Then according to post-processing and denormalization function $ \Phi^{-1}\left( \cdot \right) $, the edge server will update the global model parameters according to
\begin{equation}
\label{quantity}
\mathbf{w}\!_t\!=\!\mathbf{w}\!_{t\!-\!1}\!-\!\zeta_t\Phi^{-1}\!\left(\!\frac{ \mathbf{y}_t }{\sigma\!_t\!\left| \mathbf{S}_t\right| }\!\right)\!=\! \mathbf{w}\!_{t\!-\!1}\!-\!\zeta_t\!\left[ \frac{\sum\nolimits_{n\in \mathbf{S}_t}\!\mathbf{g}_{n,t}}{\left|\mathbf{S}_t\right|}\!+\!\hat{\bm{\varepsilon}}\!_t\right]\!,
\end{equation}
where $\hat{\bm{\varepsilon}}_t \!=\! \Phi^{-1}\left(\frac{\bm{\varepsilon}_t}{\sigma_t\left| \mathbf{S}_t\right|} \right) $ and $ \zeta_t $ is the learning rate in the $ t_{th} $ communication round.
Here, we assume that all edge devices can receive the global model broadcasted by the edge server without error,
considering that the edge server has sufficient power to resist the channel fading and noise.
Then according to (\ref{equ:y_t}),
the received SNR at the server in round $t$ is
\begin{equation}
\label{gamma}
\gamma_s =\frac{\sigma_t^2E\left[ \left\| \mathbf{x}_{n,t}\right\|_2^2\right] }{\sigma_0^2}.
\end{equation}
Let $\gamma_\text{thr}$ denote the pre-defined SNR threshold, which requires that $\gamma_s \geq \gamma_\text{thr}$ at the edge server.
Then $ \sigma_t^2 $ can be obtained by
\begin{equation}
\sigma_t^2 =\gamma_\text{thr}\sigma_0^2.
\label{equ:sigma}
\end{equation}
\subsection{AirComp enabled Federated Learning Mechanism}
\vspace{-2mm}
\label{model_learning}

\begin{figure}[!t]
	\centering
	\includegraphics[width=0.498\textwidth]{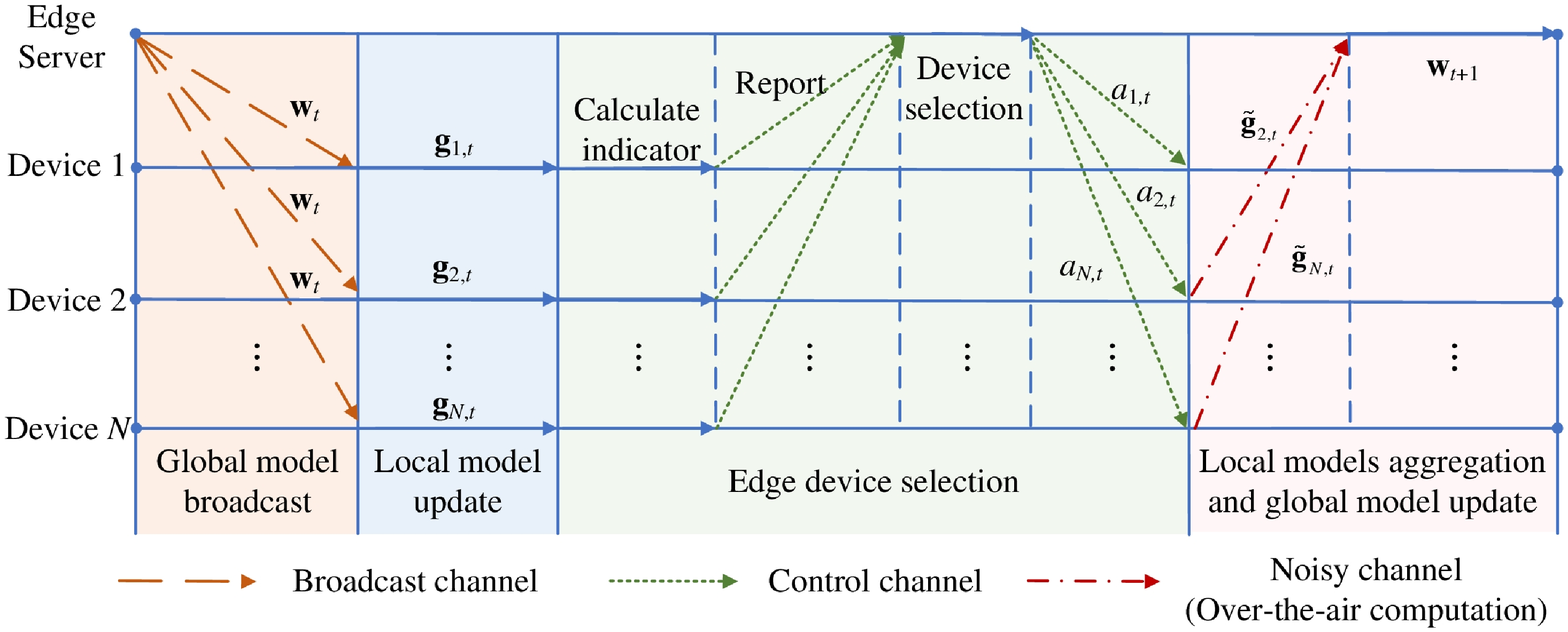}\vspace{-2mm}
	\caption{Gradient and channel aware scheduling policy in FEEL systems.}\vspace{-3mm}
	\label{fig:FL}
\end{figure}

As shown in Fig.~\ref{fig:FL}, the edge server and devices in the FEEL system are operated to implement the channel-and-data quality aware process as the following federated training steps.

\subsubsection{Global Model Broadcast}
At the beginning of communication round $t$, the edge server broadcasts the current global model $ \mathbf{w}_t $ to all edge devices for their local model training.

\subsubsection{Local Model Update}

After receiving global model $ \mathbf{w}_t $, each device computes the gradient of local model by running an SGD algorithm~\cite{fl} on a local mini-batch $ \mathcal{D}_{n,t}^m \subseteq \mathcal{D}_{n}  $, according to\footnote{In this work, all devices need to compute the gradient since that the computation results will be required in the following device selection strategy to improve the training efficiency.
In addition, the gradients of unselected devices will be utilized in subsequent rounds when the device qualities are satisfied.
We will introduce these detailed mechanisms in Sections~\ref{artm} and~\ref{device quality indicator}.}	
\begin{equation}
\mathbf{g}_{n,t} = \frac{1}{\left|\mathcal{D}_{n,t}^m\right|} \sum\nolimits_{\left( \mathbf{s}_{i}, q_{i} \right)\in \mathcal{D}_{n,t}^m} \nabla f_{n}\left( \mathbf{w}_{t-1};\mathbf{s}_{i}, q_{i}\right),
\label{equ:local_update}
\end{equation}
where local mini-batch $ \mathcal{D}_{n,t}^m $ is a subset of training samples and labels selected uniformly and randomly from dataset $\mathcal{D}_{n}$ in communication round $t$ to reduce the complexity of training.

\subsubsection{Edge Device Selection}

In order to further reduce the latency and resource overhead, only parts of edge devices in the system are selected to transmit their gradient vectors to the server for aggregation in one communication round.
These devices are selected according to their expected contribution on improving the accuracy and convergence of global model, which depends on the quantity of them and their ``qualities''.
To be specific, (\ref{quantity}) indicates that the quantity of edge devices selected, i.e., $\left|\mathbf{S}_t\right|$, can ensure robust received signals against the channel noise.
On the other hand, the ``qualities'' of device candidates need to be measured by their channel gains, significance of their local updates and transmit energy consumptions, which can ensure that the limited available energy of these devices can be used efficiently only when their local updates are important and their channel qualities to the server are qualified and sufficient for reliable transmissions.

As shown in Fig.~\ref{fig:FL}, at the beginning of this step, each edge device $n\in \mathsf{\mathcal{N}}$ calculates its data state information (DSI) indicator $ v_{DSI_{n,t}}$ and CSI indicator $ v_{CSI_{n,t}}  $ according to its gradient ${\left\| \tilde{\mathbf{g}}_{n,t}\right\| _2^2}$ and channel gain ${\left| h_{n,t}\right|}$, respectively, which can measure the significance of local update and channel quality of device $n$ in round $t$.
Along with its energy consumption $E_{n,t}$, device $n$ generates a report of its local status denoted by $\left( v_{DSI_{n,t}}, v_{CSI_{n,t}},E_{n,t} \right)$, and then submits it to the server.
Based on these received reports from all devices, the server makes the decision of device selection, denoted by $\left\{ a_{n,t}, \forall n\right\}$, by applying an appropriate mechanism.
Such a device selection
mechanism will be proposed in Section~\ref{mechanism} to achieve a high-accuracy and energy-efficient gradient aggregation for global model training at the edge server.

\subsubsection{Local Models Aggregation and Global Model Update}

In this step, the selected edge devices will transmit their local model parameters to the edge server via the AirComp scheme to realize the local model aggregation.
We note that there exist the unselected devices, the gradient vectors of which do not participate in the model updating of current round.
To retain these valid gradient vectors of unselected devices and ensure them play their roles in the future updating, $\mathbf{g}_{n,t}$ ($\forall n \notin  \mathbf{S}_t$) will be accumulated as residuals $\mathbf{r}_{n,t}$ to seize the future opportunity of transmission instead of being discarded roughly.
Therefore, each selected device $n\in  \mathbf{S}_t$ uploads the combination of current gradient vector and accumulated residual, denoted by $ \tilde{\mathbf{g}}_{n,t} =\mathbf{g}_{n,t}+\mathbf{r}_{n,t}$, as its local update information to upload to the edge server via AirComp. The detailed residual operation will be explained in Section~\ref{residual}.
After aggregating the local update information from all selected devices, the edge server updates the global model according to (\ref{quantity}) for the next communication round.

\section{Problem Formulation of Accumulated Residual based Federated Learning}
\label{residual}

To take advantage of the local gradients from the unselected devices and then accelerate the convergence rate,
we will design and introduce an accumulated residual based mechanism for local model uploading and aggregation in Section~\ref{artm}, and then formulate the optimization problem of device selection at the server in Section~\ref{convergence analysis}.
In addition, we analyze the convergence performance of the designed FEEL system in Section~\ref{convergence_assumption}, based on which we transform the optimization problem formulated in Section~\ref{convergence analysis} into a tractable form.

\subsection{Accumulated Residual Transmission Mechanism}
\label{artm}

Resulting from the device selection mechanism, the local data of unselected devices is not utilized sufficiently for global model updating.
This problem is not realized in most of current studies, such as~\cite{scheduling3, scheduling7, scheduling5, scheduling1, scheduling4, scheduling6}, in which the gradients trained by the unselected devices are directly discarded.
However, such a rough operation may result in training bias and change the direction and magnitude of the gradient, and further affect the stability and accuracy of model training~\cite{scheduling2,TON2022DU}.
Therefore, it is necessary to preserve some effective gradients of unselected devices and accumulate them to the further gradient updating, and then transmit the accumulated gradients to the server when the data and channel qualities are both satisfied.
For this purpose, we propose a dynamic residual feedback scheme, which allows devices to transmit past local gradients not transmitted to the edge server yet.
Specifically, the unselected devices preserve their current gradients locally and wait for further transmission.
Define that the local update information of each selected edge device $n$, denoted by $\tilde{\mathbf{g}}_{n,t}$,
is the combination of gradient vector $\mathbf{g}_{n,t}$ of current communication round and accumulated residual $\mathbf{r}_{n,t}$ of previous round. Then $\tilde{\mathbf{g}}_{n,t}$ and $\mathbf{r}_{n,t}$ are given by
\begin{subequations}
	\label{equ:residual}
	\begin{align}
	&\tilde{\mathbf{g}}_{n,t}=\mathbf{g}_{n,t}+\mathbf{r}_{n,t},  \label{equ:tilde_g}\\
	& \mathbf{r}_{n,1}=\mathbf{0},~\mathbf{r}_{n,t}=\begin{cases}\mathbf{0},&n\in \mathbf{S}_{t-1},~t\ge 2,\\ \xi\mathbf{g}_{n,t-1},&n\notin \mathbf{S}_{t-1},~t\ge 2,\end{cases} \label{equ:r}
	\end{align}
\end{subequations}
respectively, where $0\le\xi\le1$ reflects the influence of previous updates on current training. Without loss of generality, we set $\xi=1$ in this work.
\begin{remark}
The residual feedback mechanism is designed to take full advantage of those local updates which should be utilized at the edge server but were not transmitted due to the energy constraint in the previous rounds. 
In other words, the residual feedback mechanism can improve the training performance to achieve the optimal benchmark, which can be obtained when the server updates the global model using the average of all noiseless local gradients, and meet the energy requirement as well.\end{remark}

\subsection{Problem Formulation}
\label{convergence analysis}

According to~(\ref{obj}), the main objective of a FEEL system is to minimize the expectation of final training loss to attain a desired accuracy of the model by optimizing the device selection strategy $a_{n,t}$. 
Applying the law of telescoping sums, the optimization problem of device selection is formulated as
\begin{subequations}
	\label{equ:p1}
	\begin{align}
	P_1: \quad &&\min_{{\left\{a_{n,t}\right\}}_{t=1}^T}~~ &\sum_{t=1}^{T}\left(E\left[ f_{\mathbf{w}_{t}}\right]-E\left[ f_{\mathbf{w}_{t-1}}\right]\right) ,&\label{equ:opt:FCP:obj}  \\
	&& \text{s}\text{.t}\text{.}~~&\frac{1}{T}\sum_{t=1}^{T}a_{n,t}E_{n,t}\leq\bar{E}_n, &\label{equ:p1:con1} \\
	&& &a_{n,t}\in\left\lbrace 0,1\right\rbrace , &\label{equ:p1:con2}
	\end{align}
\end{subequations}
where the constraint~(\ref{equ:p1:con1}) represents the average transmit energy constraint $ \bar{E}_n $ on device $n$ over $T$ rounds.
\begin{remark}
According to the form of problem $P_1$ in (\ref{equ:p1}), it is difficult to express an explicit form of $ E\left[ f\left( \mathbf{w}_t\right) \right]-E\left[ f\left( \mathbf{w}_{t-1}\right) \right]  $ due to the unpredictable model parameters in neural network architectures.
To solve problem $P_1$, it is necessary to substitute it with its upper bound
in an approximate closed-form
based on its convergence analysis, which will be
provided in the following subsection.
\end{remark}

\subsection{Convergence Analysis and Problem Transformation}
\label{convergence_assumption}

We first introduce the following assumptions provided in the state-of-the-art studies~\cite{scheduling2, scheduling3, scheduling1, scheduling6}:


\textit{1) Loss functions $ f_n(\mathbf{w}), \cdots , f_N(\mathbf{w}) $ are $L$-smooth, i.e., $ \forall \mathbf{x}, \mathbf{y} \in \mathcal{R}^{L_g} $ and $\forall n \in \mathcal{N} $, $\exists L<\infty$, it satisfies}
\begin{equation}
f_n(\mathbf{y})\leq f_n(\mathbf{x})+\nabla f_n(\mathbf{x})^T(\mathbf{y}-\mathbf{x})+\frac{L}{2}\left\| \mathbf{y}-\mathbf{x}\right\| ^2 .
\end{equation}

\textit{2) Loss functions  $ f_1(\mathbf{w}), \cdots , f_N(\mathbf{w}) $ are $\mu$-strongly convex, i.e., $ \forall \mathbf{x}, \mathbf{y} \in \mathcal{R}^{L_g} $ and $ \forall n \in \mathcal{N} $, it satisfies}
\begin{equation}
f_n(\mathbf{y})\geq f_n(\mathbf{x})+\nabla f_n(\mathbf{x})^T(\mathbf{y}-\mathbf{x})+\frac{\mu}{2}\left\| \mathbf{y}-\mathbf{x}\right\| ^2  .
\end{equation}

\textit{3) Stochastic gradient is unbiased and variance bounded, i.e., $ \forall n \in \mathcal{N} $ and $ \forall t \in \mathcal{T} $, taking the expectation over stochastic data sampling, it satisfies}
\begin{subequations}
\begin{align}
E\left[\mathbf{g}_{n,t}\right]\left|_{\mathcal{D}_{n,t}^m \subseteq \mathcal{D}_n}\right.&=\mathbf{g}_t ,\\
E\left[\left\| \mathbf{g}_{n,t}-\mathbf{g}_t\right\| _2^2\right]\left|_{ \left(\mathbf{s}_{i},q_{i} \right)\in \mathcal{D}_n}\right. &\leq G^2, \forall n \in \mathcal{N} ,
\end{align}
\end{subequations}
\textit{where
constant $G^2$ denotes the variance of local gradients and $ \mathbf{g}_{t} $ is the global model gradient in round $ t $, $\forall n,t$}.

%

Based on these assumptions above, we first provide the expectation of average residual in each communication round in Lemma~\ref{lemma1} for further convergence analysis.

\begin{lemma}
	\label{lemma1}
	Given global model gradient $ \mathbf{g}_{t} $ and variance of local gradients $G^2$ in round $t$, the upper bound of average accumulated residual in each round is
	\begin{eqnarray}
	\begin{aligned}
	E\left[ \frac{1}{N}\sum_{n=1}^{N}\mathbf{r}_{n,t+1}\right] \leq \frac{Pr_t\left( G^{2}+\left\|  \mathbf{g}_{t} \right\| _{2}^{2}\right)}{1 - Pr_t},
	\end{aligned}
	\label{equ:residual bound}
	\end{eqnarray}
where $ 0< Pr_t < 1 $ denotes the average possibility of that the device is unselected in round $t$, $\forall t \in \mathcal{T} = \left\lbrace 0,\cdots, T-1 \right\rbrace$.
\end{lemma}

Proof: See Appendix~\ref{proof:lemma1}.

Next, we provide the one-round convergence analysis by characterizing the upper bound of $ E\left[ f\left( \mathbf{w}_t\right) \right]-E\left[ f\left( \mathbf{w}_{t-1}\right) \right] $ in the following Lemma~\ref{lemma2}, based on assumptions and Lemma~\ref{lemma1} above.
Here we consider the transmit power, channel noise, residual upper bound as well as the stochastic gradients jointly in the AirComp enabled FEEL system.

\begin{lemma}
	\label{lemma2}
	Given a set of selected devices $\mathbf{S}_{t}$ and mini-batch $ \mathcal{D}_{n,t}^m$ in round $t$, the upper bound of one-round convergence is
	\begin{eqnarray}
	\begin{aligned}
	&E\left[ f\left( \mathbf{w}_{t} \right)\right] - E\left[ f\left( \mathbf{w}_{t - 1} \right) \right] \leq \frac{L\eta_{t}^{2}}{2} \left\| \mathbf{g}_{t}\right\|_2^2+\frac{L\eta_{t}^{2}+\eta_t}{2}m^2\\
	&+\!\frac{L\eta_{t}^{2}\!+\!\eta_t}{2}\!\frac{Pr_t\!\left(\! G^{2}\!+\!\left\|  \mathbf{g}_{t} \right\| _{2}^{2}\!\right)}{1 \!-\! Pr_t}\!+\!\frac{L\eta_{t}^2}{2}\!\left[ \!\frac{\delta^2}{\gamma_{thr} \left| \mathbf{S}_{t} \right|^{2}}\!+\!\frac{G^2}{\left| \mathbf{S}_{t} \right|\left|\mathcal{D}_{n,t}^m\right|}\!\right]\!.
	\end{aligned}
	\label{equ:convergence upper bound}
	\end{eqnarray}			
\end{lemma}
where $\delta^2$ and $m$ is the estimate variance and mean of the global update information, respectively.

Proof: See Appendix~\ref{proof:lemma2}.

According to Lemma 2, we note that the quantity of selected devices, i.e., $ \left| \mathbf{S}_t\right| = \sum_{n\in \mathbf{S}_t} a_{n,t} $,
plays an important role in the training performance to resist the impact of channel noise. Therefore, the goal of our efficient device scheduling mechanism is to optimize selection strategies $ a_{n,t} $ ($\forall n, t$)
according to the channel noise level in each round.

Instead of maximizing the average number of selected devices per round as studied in~\cite{mse, maxnum}, Lemma~\ref{lemma2} provides us with a reasonable direction to minimize the upper bound of $ E\left[ f\left( \mathbf{w}_t\right) \right]-E\left[ f\left( \mathbf{w}_{t-1}\right) \right] $ alternatively to resist the channel fading and noise.
Such an online operation is executable since the future state of the system is unavailable.
Moreover, $ \mathbf{g}_t $ is the global gradient defined on the whole dataset, which indicates that $ \mathbf{g}_t $ is fixed for a given dataset.
Then combined with hyper-parameters including learning rate $ \zeta_t $, smoothness parameter $ L $, estimate mean of global update information $m$ and $ G^2 $,
the first three terms of~(\ref{equ:convergence upper bound}) formulated in Lemma~\ref{lemma2} are constants.
Therefore, ignore constant ${L{\zeta_{t}^{2}}}/{2}$ and let\vspace{-1mm}
\begin{eqnarray}
	U_t\triangleq \frac{\delta^2}{\gamma_{\text{thr}} \left| \mathbf{S}_{t} \right|^{2}}+\frac{G^2}{\left| \mathbf{S}_t\right| \left|\mathcal{D}_{n,t}^m\right|} ,
	\label{equ:U}
\end{eqnarray}
and then optimization problem $P_1$ can be transformed as
\vspace{-1mm}
%
\begin{subequations}
	\label{equ:P2}
	\begin{align}
	P_2:\quad && \min~~ & \sum_{t=1}^{T}U_t,&  \label{equ:P2:ob}\\
	&& \text{s}\text{.t}\text{.}~~ & \frac{1}{T}\sum_{t=1}^{T}a_{n,t}E_{n,t}\leq\bar{E}_n, & \\
	&& &a_{n,t}\in\left\lbrace 0,1\right\rbrace .&
	\end{align}
\end{subequations}

\section{Gradient and Channel Aware Device Scheduling for AirComp}
\label{mechanism}

This section will propose a gradient and channel aware dynamic scheduling mechanism for AirComp enabled FEEL systems.
As introduced in Section~\ref{model_learning},
the data qualities, channel conditions, as well as energy consumptions of edge devices are considered by the edge server to decide which devices are qualified for model aggregation.
So we first design the indicator of device quality considering these criteria above to select satisfied devices in Section~\ref{device quality indicator}.
Based on this defined indicator and the optimization problem $P2$ formulated as~(\ref{equ:P2}) in Section~\ref{residual}, we adopt the Lyapunov-drift-optimization-based approach to formulate the dynamic device selection problem in Section~\ref{problem formulation}, and then the algorithm of this problem is
designed in Section~\ref{algorithm}.

\subsection{Indicator of Device Quality}
\label{device quality indicator}

In most current studies, devices are selected randomly to participate the model aggregation in federated learning, which ignores the differences between devices and channel conditions~\cite{scheduling1, scheduling4}.
However, updating local model parameters that not vary much will cause the waste and conflict of transmission resource, while with limited performance improvement of global model training. Additionally, both energy consumption of devices for transmitting their local updates and their channel conditions to the server can affect the efficiency and accuracy of model aggregation.
Therefore, it is important to design an appropriate indicator of device quality, considering the factors above, to measure their potential contributions to the model trailing in the FEEL system.
To be specific, the significance of local update information is measured by its $ l_2$-norm, i.e., ${\left\| \tilde{\mathbf{g}}_{n,t}\right\| _2^2}$.
We consider that devices with larger values of ${\left\| \tilde{\mathbf{g}}_{n,t}\right\| _2^2}$ will contribute more to the global model training than those with smaller values~\cite{scheduling6}. In addition, we use $ \left| h_{n,t}\right|$ to quantify the channel condition, i.e. CSI, and a larger value of $ \left| h_{n,t}\right|  $ means a better channel condition~\cite{scheduling7}.
Then the indicators DSI and CSI with unified value, i.e., $ v_{DSI_{n,t}}$ and $v_{CSI_{n,t}}$ mentioned in Section~\ref{model_learning}, can be defined as
\begin{equation}
\label{19}
v_{DSI_{n,t}}=\dfrac{\left\| \tilde{\mathbf{g}}_{n,t}\right\| _2^2}{g_{max}},~~v_{CSI_{n,t}}=\dfrac{\left|  h_{n,t}\right|}{h_{max}},
\end{equation}
respectively, where
$g_{max}$ is the maximum of $\left\| \tilde{\mathbf{g}}_{n,t}\right\| _2^2$,
and $h_{max}$ is the maximum of $\left| h_{n,t}\right|$, $\forall n\!\in \!\mathsf{\mathcal{N}}$, $\forall t\!\in \!\mathsf{\mathcal{T}}$.
Based on the definitions above, the importance of device $n$ considering both its update and channel condition can be defined as
\begin{equation}
V_{n,t}=\rho_{1}v_{DSI_{n,t}}+\rho_{2}v_{CSI_{n,t}},
\label{equ:V}
\end{equation}
where weight parameters $\rho_{1}$ and $\rho_{2}$ jointly balance the device quality of DSI and CSI, and $\rho_{1} + \rho_{2}=1$.


Compared with local training, communications to the server cause more energy consumption for devices.
So for an energy-efficient FEEL system, the energy consumption of different devices for transmission should be also considered when designing device selection mechanisms~\cite{indicator}.
According to the channel inversion based power control policy formulated in (\ref{equ:p}),
the transmit energy of device $ n $ is given by $ E_{n,t}\!=\!{\sigma_t^2}/{\left| h_{n,t}\right| ^2} $.
Based on the definitions above, the optimization problem of device selection considering the data quality, channel condition and energy consumption is formulated as
\vspace{-3mm}
\begin{subequations}
	\label{equ:P3}
	\begin{align}
	P_3:\quad && \min~~ & \lambda_E\sum_{n=1}^{N}a_{n,t}E_{n,t}\!-\!\lambda_V\sum_{n=1}^{N}a_{n,t}V_{n,t},&\label{equ:P3:ob}\\
	&& \text{s}\text{.t}\text{.}~~ & \frac{1}{T}\sum_{t=1}^{T}a_{n,t}E_{n,t}\leq\bar{E}_n,& \label{equ:P6:con1} \\
	&& &a_{n,t}\in\left\lbrace 0,1\right\rbrace ,& \label{equ:P6:con2}
	\end{align}
\end{subequations}
where $ \lambda_E $, $ \lambda_V>0 $ are adjustable weight parameters to balance the loss of energy consumption and the gain of update and channel qualities. Without loss of generality, let $\lambda_E+\lambda_V=1$.

We define the comprehensive indicator of device quality considering the device data, channel and energy consumption, which can be given by
\vspace{-1mm}
\begin{equation}
I_{n,t}\!=\!\lambda_VV_{n,t}\!-\!\lambda_EE_{n,t}. 
\end{equation}
According to (\ref{19}) and the designed accumulated residual transmission mechanism in Section~\ref{artm}, one can notice that the calculation of $ I_{n,t} $ should use the local updates including the current updates and previous updates which have not been transmitted.
Furthermore, the value of $ I_{n,t} $ can be calculated locally at each device and then transmitted to the edge server for device selection by solving $P_3$, which can avoid the privacy disclosure of local datasets.

\vspace{-2.5mm}
\subsection{Lyapunov-Drift-Minimization Based Problem Formulation}
\label{problem formulation}
\vspace{-1mm}

As formulated in (\ref{equ:P2:ob}) and (\ref{equ:P3:ob}),
the quantity and quality of selected devices are two key factors in device selection, which can be optimized by solving $P_2$ and $P_3$, respectively.
Based on problems $P_2$ and $P_3$, we aim to design a scheduling mechanism which
can decide the quantity of selected devices to reduce the impact of channel noise in each communication round effectively, and ensure that the quality of selected devices is as high as possible to take the full advantage of their training performances as well.
Such a scheduling mechanism can be considered as a multi-objective joint optimization problem to obtain the linear tradeoff between the maximization of device quality and minimization of the objective function in FEEL systems, which can be solved by the drift-plus-penalty algorithm in Lyapunov optimization~\cite{lyapunov}.
Therefore, we formulate the optimization problem of the device scheduling mechanism as
\vspace{-2mm}
\begin{subequations}
	\label{equ:P4}
	\begin{align}
	P_4:\quad  && \begin{aligned}&\underset{a_{n,t}}\min\end{aligned}	&\alpha U_t-\sum_{n=1}^{N}a_{n,t}I_{n,t} , & \label{equ:opt:FCP:obj4}\\
	&& \text{s}\text{.t}\text{.}~~&\frac{1}{T}\sum_{t=1}^{T}a_{n,t}E_{n,t}\leq\bar{E}_n,& \label{equ:opt:FCP:con1} \\
	&& & a_{n,t}\in\left\lbrace 0,1\right\rbrace, &\label{equ:opt:FCP:con2}
	\end{align}
\end{subequations}
where $ \alpha $ is the Lyapunov factor to balance the loss of $U_t$ defined in (\ref{equ:U}) and device quality $I_{n,t}=\lambda_VV_{n,t}-\lambda_EE_{n,t}$.

According to (\ref{equ:P4}), the individual energy consumptions are considered in the proposed mechanism. However, the proposed mechanism can be well applied to sum-energy-constraint-based FEEL system as well, since a proper setting of $ \lambda_E $, $ \lambda_V $ and $ \alpha $ can enable an acceptable sum energy consumption in a single round by distributing the limited power to energy-efficient devices.


%

\subsection{Dynamic Device Scheduling Algorithm}
\label{algorithm}

\begin{algorithm}[!t]
 \caption{Gradient and Channel Aware Device Scheduling}
  \label{alg:1}
 \begin{algorithmic}[1]  
  \Require $ \lambda_E $, $ \lambda_V $, $ \alpha $, $ \gamma_{thr} $.
  \Ensure  The device scheduling mechanism $a_{n,t}$.
  \State Initialization $ \mathbf{w}_0 $.\label{alg:step01}
  \For {$t=1 \to t=T$}
  \State The server broadcasts global model $ \mathbf{w}_t $.
  \For {$n=1 \to n=N$}
  \State Local model update according to (\ref{equ:local_update}) for $ \mathbf{g}_{n,t} $.
  \State Calculate $ I_{n,t}  $ according to (\ref{equ:vt}) and
  \Statex $ \quad \qquad $transmit to the server via control channel.
  \EndFor \label{alg:step07}
  \State The server sorts $ I_{n,t} $ in descending order \label{alg:step1}
  \For {$k=1 \to k=N$} \label{alg:step2}
  \State The server calculates $ p_t\left( k\right) $ according to
  \begin{equation}
     p_t\left( k\right)=\alpha U_t-\sum_{n=1}^{k}I_{n,t}.\label{equ:vt}
  \end{equation}
  \EndFor \label{alg:step4}
  \State $k^*=arg \min p_t\left( k\right) $ \label{alg:step5}
  \For {$n=1 \to n=N$} \label{alg:step6}
  \If{$I_{n,t}\geq I_{k^*,t} $}
  \State $a_{n,t} = 1$
  \Else
  \State $a_{n,t} = 0$
  \EndIf
  \EndFor \label{alg:step10}
  \For {$n=1 \to n=N$}
  \State Update local residual according to (\ref{equ:r}).
  \If{$a_{n,t} = 1$}
  \State Set transmit power according to (\ref{equ:p});
  \State Transmit $ \tilde{\mathbf{g}}_{n,t} $ according to (\ref{equ:tilde_g}) to the server.
  \EndIf
  \EndFor
  \State Model aggregation and global model update by (\ref{quantity}).
  \EndFor
 \end{algorithmic}
\end{algorithm}

This section proposes a dynamic device scheduling algorithm based on Lyapunov optimization to achieve device selection, which is summarized in Algorithm~\ref{alg:1} in detail.

In Steps~\ref{alg:step01} to~\ref{alg:step07}, the edge server and devices complete the \emph{Global Model Broadcast} and \emph{Local Model Update} steps designed in Section~\ref{model_learning}, respectively, and indicator $I_{n,t}$ is calculated and then reported to the server by each device $n\in \mathsf{\mathcal{N}}$ in each round $ t\in \mathsf{\mathcal{T}}$.
In Step~\ref{alg:step1}, the server sorts $ I_{n,t}, \forall n\in \mathsf{\mathcal{N}}$ in descending order by a quick sorting algorithm.
A large value of $ I_{n,t} $ means good device quality and has high priority of being selected.
In Steps~\ref{alg:step2} to~\ref{alg:step4}, the server iterates over the possible quantity of selected devices from $ k=1 $ to $ k = N $, and calculates the corresponding drift-plus-penalty denoted by $ p_t\left( k\right) $.
In Step~\ref{alg:step5}, $ k^* $ is the optimal quantity of selected devices corresponding to the minimum $ p_t\left( k\right) $.
Then the devices with $ k^* $ largest values of $ I_{n,t} $ will be selected in round $ t $, as shown in Steps~\ref{alg:step6} to~\ref{alg:step10}.
In addition, the complexity of the algorithm summarized in Steps~\ref{alg:step1} to~\ref{alg:step10} is $ o \left( N \right)  $, hence the complexity of Algorithm~\ref{alg:1} in a single round is $ o\left( NlogN\right)  $.
\begin{remark}
Although transmitting $ I_{n,t} $ will bring extra overheads, such a transmission cost can be negligible since the amount of data $ I_{n,t} $ is far less than that of local updates. 
In addition, the proposed device scheduling mechanism is flexible and extensible by substituting and re-defining $ I_{n,t} $ with different measurements.
Furthermore, the proposed mechanism can be well applied to situations with non-i.i.d. datasets, which will be validated in Section~\ref{non_iid}.
\end{remark}


\subsection{Performance Analysis}
\label{convergence}

This section analyzes the convergence performance of the proposed dynamic device scheduling policy based on Lyapunov drift and dynamic residual.
Here we define the Lyapunov function as $ L_t=f\left( \mathbf{w}_t\right) -f^* $, where $ f $ is the loss function, and $ f^* $ is the optimal value of $ f $. Let the $T$-round Lyapunov drift term as $ \Delta_{T} =E\left[ L_{1\!+\!T}-L_{1} \right]$.
Then we provide the following Theorem~\ref{theorem1} to illustrate the convergence performance of the proposed mechanism.

\begin{theorem}
	\label{theorem1}
	Given global model vector $\mathbf{w}_0$ and device scheduling set $\left\{\mathbf{S}_{t}, t\in \mathsf{\mathcal{T}}\right\}$, then the $T$-round Lyapunov drift term is bounded by
    \begin{eqnarray}
    \label{26}
    	\begin{aligned}
    		&\mathrm{\Delta}_{T} = E\left[  f\left(\mathbf{w}_T \right)  \right]-f\left(\mathbf{w}_0 \right) \\
    		&\leq \left(\prod_{i=1}^{T} C_i-1\right)  \left[ E\left[ f\left(\mathbf{w}_0 \right)\right]\!-\!f^*\right]  \!+\!\sum_{i=1}^{T-1}B_i\prod_{j=i+1}^{t}C_j \!+\! B_T.
    	\end{aligned}
    \end{eqnarray}
where $B_t\!\triangleq\! \frac{L\zeta_{t}^{2}\!+\!\zeta_t}{2}m^2
	\!+\!\frac{L\zeta_{t}^{2}\!+\!\zeta_t}{2}\!\frac{Pr_tG^{2}\!}{1 \!-\! Pr_t}\!+\!\frac{L\zeta_{t}^2}{2}\left[ \!\frac{\delta^2}{\gamma_{thr} \left| \mathbf{S}_{t} \right|^{2}}\!+\!\frac{G^2}{\left| \mathbf{S}_{t} \right|\left|\mathcal{D}_{n,t}^m\right|}\!\right]$
\vspace{-1mm}
and	
$ C_t\!\triangleq\! 1 \!+\!2\mu\left(\frac{L\zeta_{t}^{2}}{2}\!+\! \frac{\left( L\zeta_{t}^{2}\!+\!\zeta_t\right) Pr_t}{2\left(1\!-\!Pr_t \right) } \right)$.
\end{theorem}

Proof: See Appendix~\ref{proof:theorem1}.
\begin{remark}
Theorem~\ref{theorem1} indicates that 
the $T$-round Lyapunov drift term, as well as the expected gap between the global training loss and its optimal value, is bounded by the three terms in (\ref{26}), which will all converge to zero as $ t $ increases.
Therefore, the fast convergence performance of the proposed algorithm is guaranteed with the communication rounds increasing.
\end{remark}
This section has proposed a communication-efficient approach to transmit sufficient and useful information under specific energy constraint and different channel conditions.
By applying the proposed mechanism, each device has an opportunity to participate the global model updating when it has a ``good'' quality in terms of its local update, energy consumption and channel condition.
Then the fairness of device selection can be also guaranteed to a certain extent.
In addition, the designed accumulated residual mechanism ensures
that the global training can utilize the distributed datasets fully to accelerate the convergence rate and accuracy of global training as well.

\section{Simulation Results}
\label{simulation}

This section conducts experiments to validate the performance of the proposed device scheduling mechanism over fading noisy channels, which are implemented on Python 3.7.

\vspace{-3mm}
\subsection{Experiment Settings}
\label{settings}
\vspace{-0.5mm}
	
We consider a FEEL system consisting of one edge server and $ N = 100 $ edge devices. The wireless channel from each device to the edge server follows the i.i.d. Rayleigh fading, modeled as $ h_{n,t} \sim CN(0,1) $ and is quasi-static within a radio resource block. Assume that the accurate channel gain can be observed by both the edge server and devices. The power scalar is calculated according to (\ref{equ:p}), where the variance of the channel noise $\sigma_0^2$ selects a value from  $\left\{0.5,1,3\right\}$ to indicate different channel conditions to validate the robustness of the proposed device scheduling mechanism.
Set the average energy budget as $ \bar{E}_n =1.5$, and the threshold of received SNR as $\gamma_\text{thr}=0\text{dB}$.

We perform the learning task of image classification by employing a convolutional neural network (CNN) with two $ 5\!\times\! 5 $ convolution layers, a fully connected layer with 512 units and ReLu activation, and a softmax output layer~\cite{fl}. We introduce the cross entropy loss function, which is smooth and strongly convex, which satisfies assumptions in Section~\ref{convergence_assumption}.
Moreover, we adopt MNIST and CIFAR-10 datasets, which consist of 60,000 and 50,000 labeled training data samples, respectively.
We consider both the i.i.d. and non-i.i.d. cases. Specifically, the i.i.d. dataset is assigned to the 100 edge devices randomly, and each device receives 600 examples (or 500 examples).
For the non-i.i.d. case, we first sort the data samples by digit labels and divide them into 200 shards with size 300 (or 250). Each device is allocated a disjoint subset of data with two shards equally. In other words, each device only has two labels~\cite{fl}.

The hyper-parameters of federated learning are set as follows. The momentum optimizer is $ 0.5 $, 
$ T=50 $ for MNIST and $ T=200 $ for CIFAR-10, the size of local mini-batch is $ \left|\mathcal{D}_{n,t}^m\right| \!=\!10$, and learning rate $\zeta_t \!= \!0.01$.
Considering that the increasing local epoch is not an effective way of improving training performance, the local epoch of local updates is set as $ 2 $, which can ensure a good convergence performance and meanwhile save the computation resource and training time.
In addition,
we set $ G^2 $ as the maximum value of all variances of local gradients in pre-training~\cite{scheduling1}.
The Lyapunov factor is set as $ \alpha \!= \!5\times10^3 $,
and let $ \rho_1\! = \!\rho_2 \!= \!0.5 $
in (\ref{equ:V})
to ensure fairness~\cite{indicator}.

\begin{figure}[!t]
	\begin{center}
		\subfigure[Quantity of selected devices]{\includegraphics[width=0.469\textwidth]{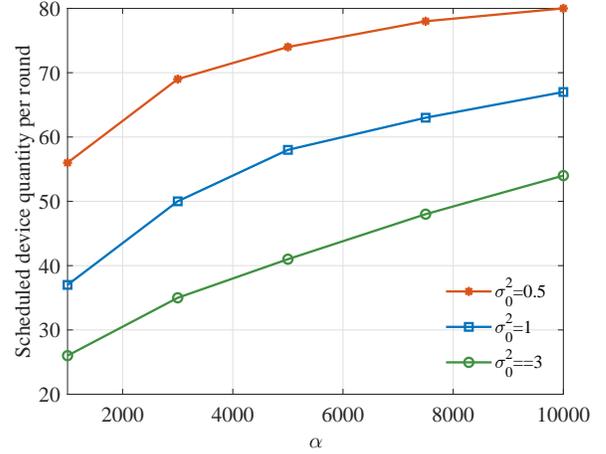}
			\label{fig:alpha_num}}
		\subfigure[Average energy consumption]{\includegraphics[width=0.469\textwidth]{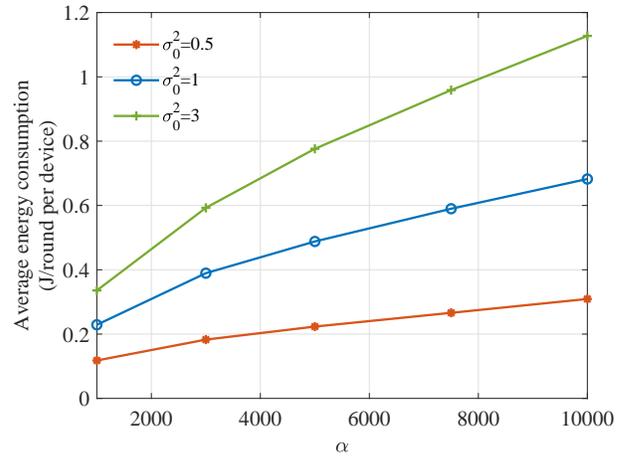}
			\label{fig:alpha_energy}}
		\caption{Quantity of selected device and average energy consumption versus different $\alpha$ and $ \sigma_0^2 $ on MNIST dataset.}\vspace{-5mm}
		\label{fig:alpha}
	\end{center}
\end{figure}

\label{alpha}
\begin{figure*}[!t]
	\begin{center}
		\subfigure[$ \sigma_0^2 $ = 0.5]{\includegraphics[width=0.31\textwidth]{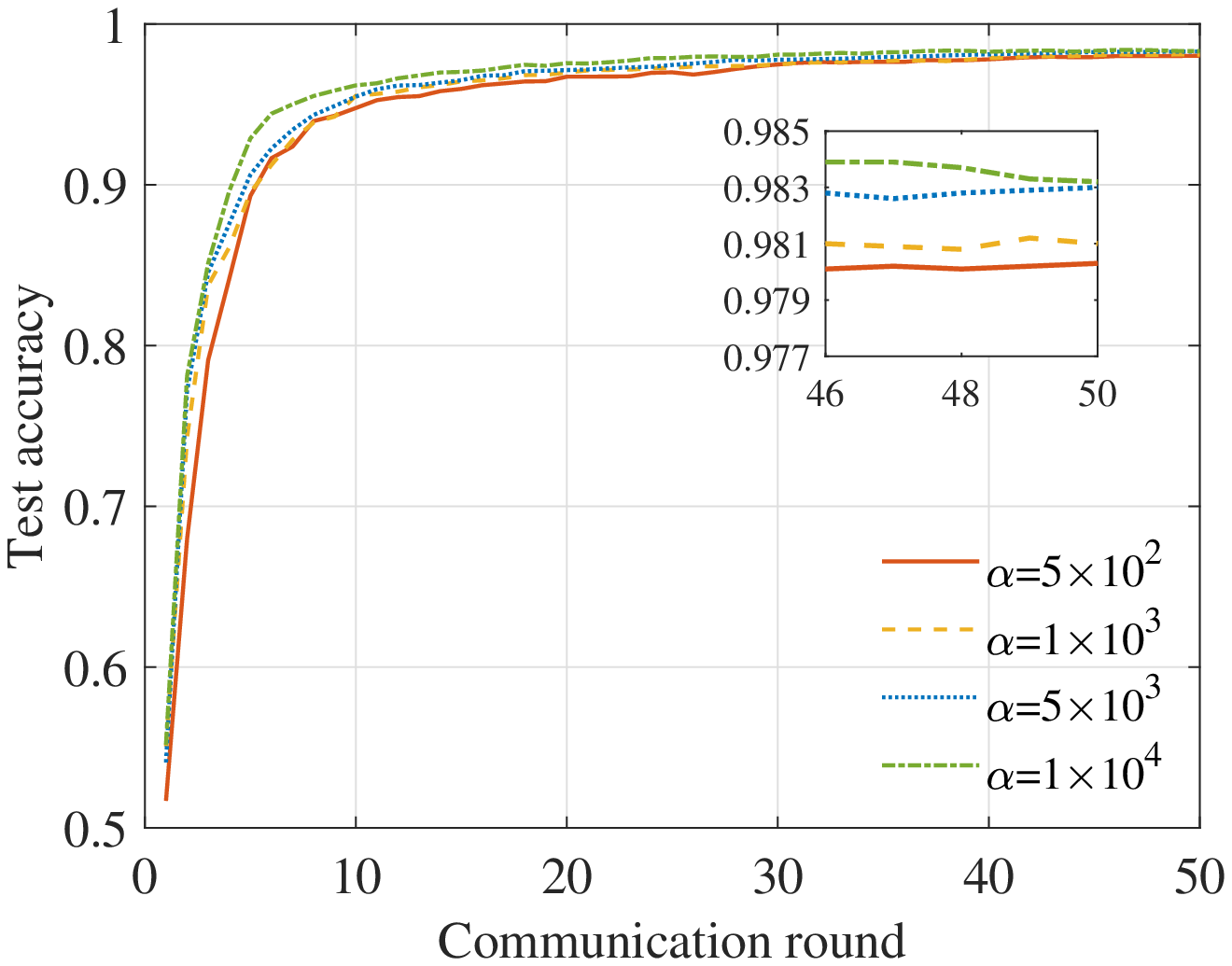}
			\label{fig:alpha_acc_05}}
		\subfigure[$ \sigma_0^2 $ = 1 ]{\includegraphics[width=0.31\textwidth]{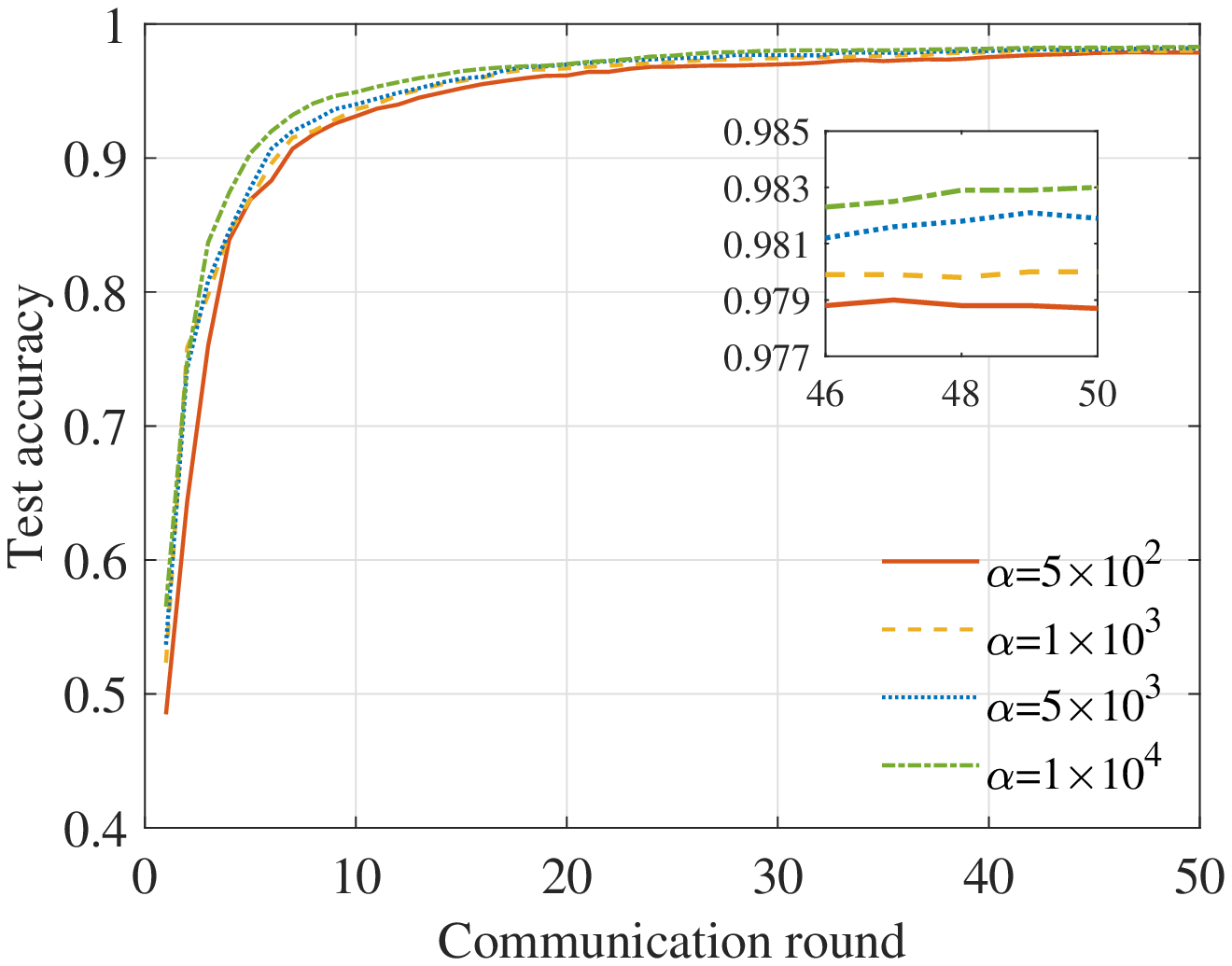}
			\label{fig:alpha_acc1}}
		\subfigure[$ \sigma_0^2 $ =3 ]{\includegraphics[width=0.31\textwidth]{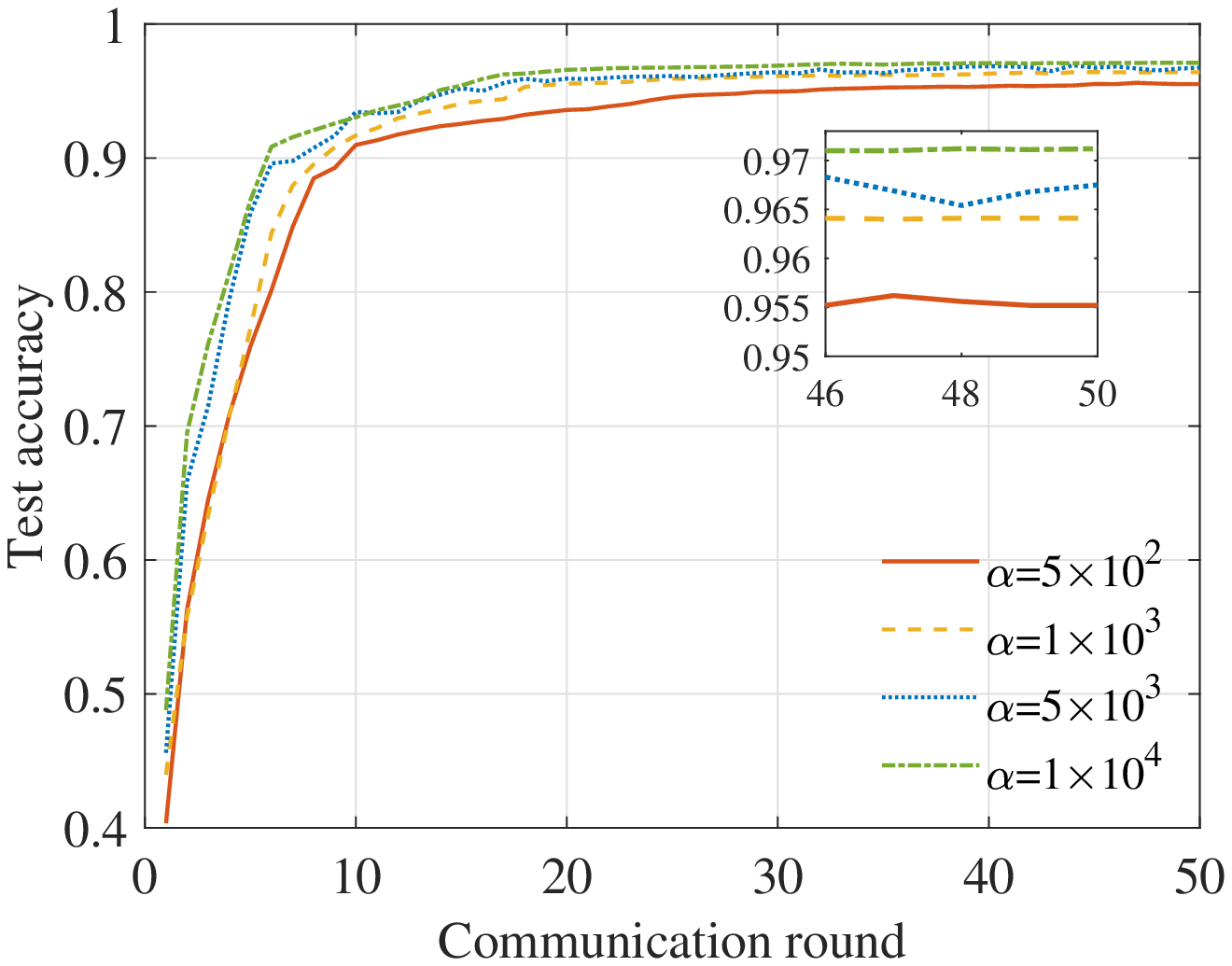}
			\label{fig:alpha_acc3}}
		\caption{Test accuracy convergency versus $\alpha$ under different channel conditions on MNIST dataset.}\vspace{-3mm}
		\label{fig:alpha_acc}
	\end{center}
\end{figure*}

\begin{figure*}[!t]
	\begin{center}
		\subfigure[Quantity of selected device]{\includegraphics[width=0.31\textwidth]{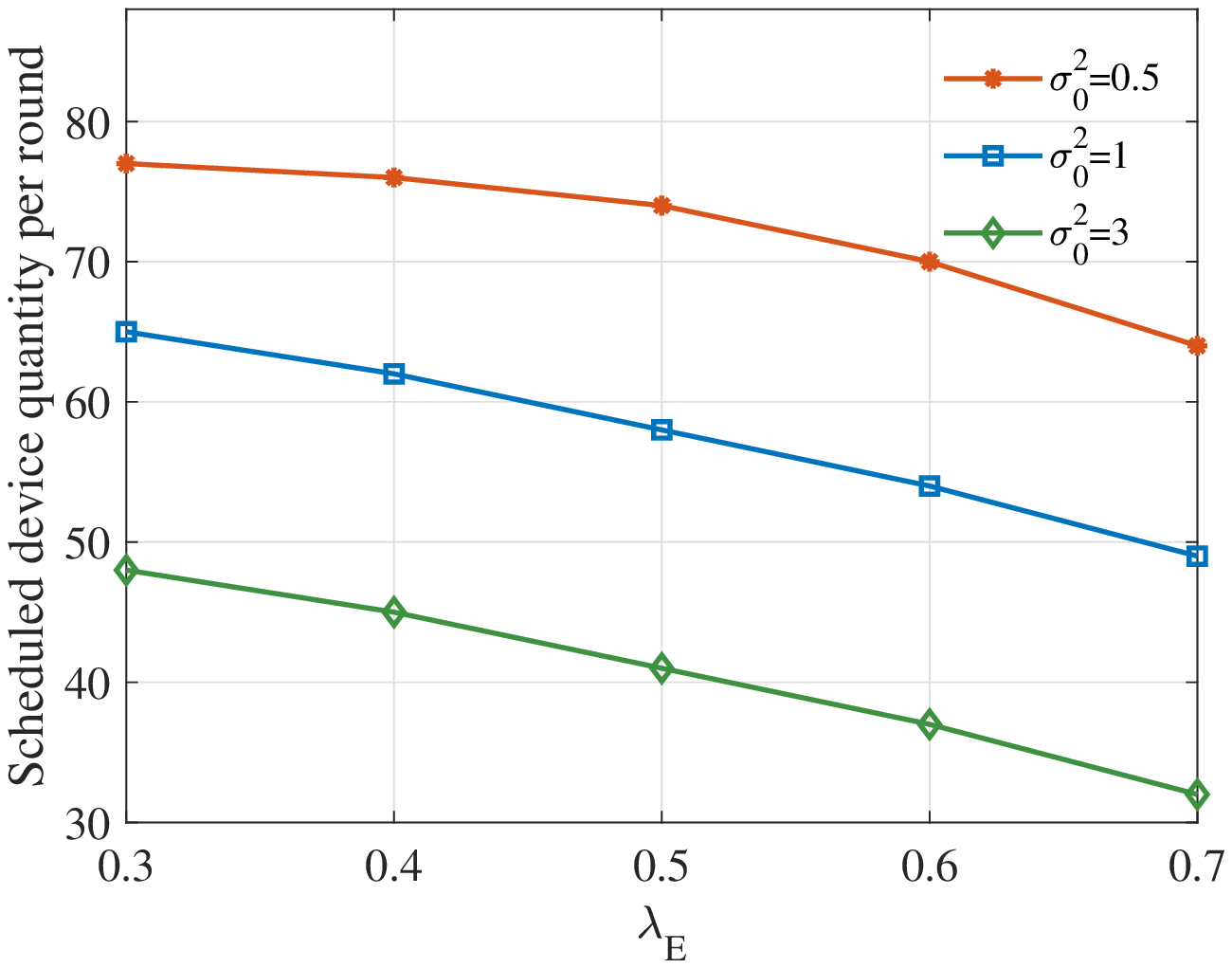}
			\label{fig:lambda_quantity}}
		\subfigure[Average energy consumption ($\sigma_0^2=1$)]{\includegraphics[width=0.31\textwidth]{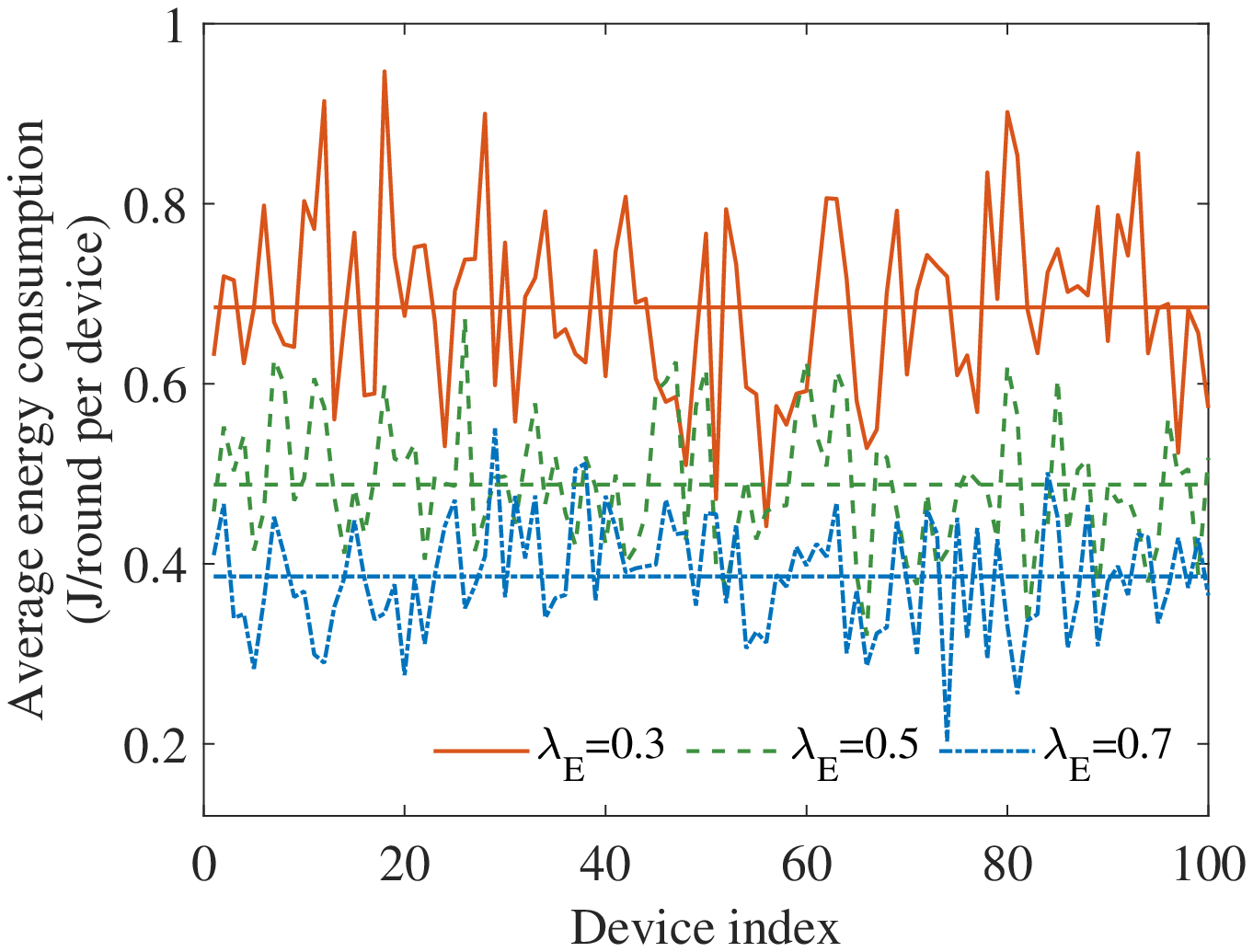}
			\label{fig:lambda_energy}}
		\subfigure[Test accuracy ($\sigma_0^2=1$)]{\includegraphics[width=0.31\textwidth]{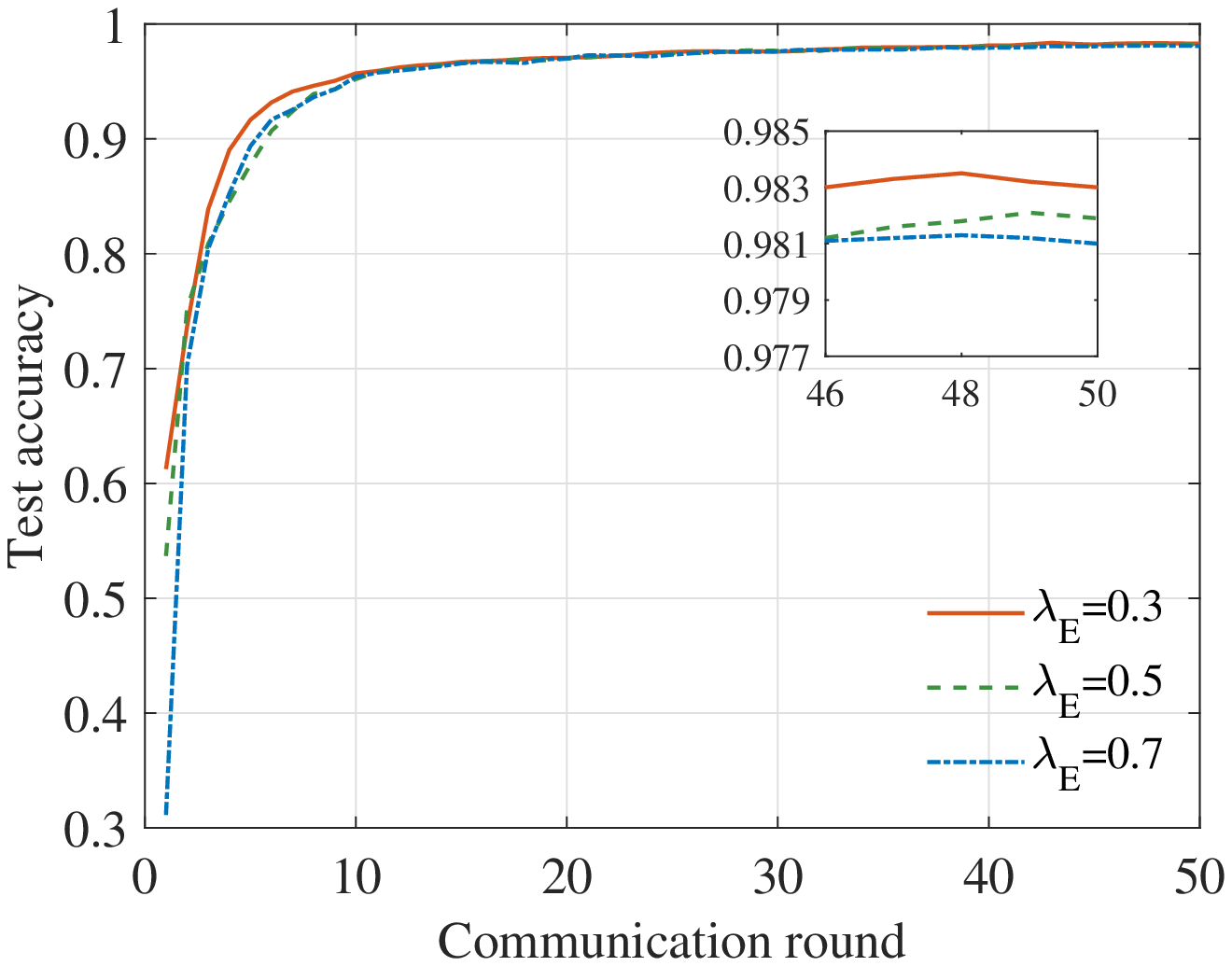}
			\label{fig:lambda_acc}}
		\caption{Quantity of selected devices, average energy consumption and test accuracy versus $ \lambda_{E} $, $ \lambda_{V} $ and $\sigma_0^2$ on MNIST dataset.}\vspace{-3mm}
		\label{fig:lambda}
	\end{center}
\end{figure*}

In this section, we compare the performance of the proposed device scheduling mechanism with the following baselines.
\begin{enumerate}
	\item \textbf{Benchmark}~\cite{fl}: The server updates the global model using the average of all noiseless local gradients, which provides the benchmark accuracy with ideal settings.
	\item \textbf{Baseline 1}~\cite{fl}: The server updates the global model by selecting $k=30$ devices randomly.
	\item\textbf{Baseline 2}~\cite{scheduling7}: In each round, device $n$ will be selected only if its channel gain satisfies $ \left| h_{n,t} \right| \geq \left| h\right|_{th} $, where $ \left| h\right|_{th} $ is the channel truncated threshold and set $ \left| h\right|_{th} \!=\! 1 $.
	\item \textbf{Baseline 3}~\cite{scheduling6}: In each round, the edge server first selects $ k_c $ devices with largest $\left| h_{n,t} \right|$, and then schedules $k$ devices with largest $\left\| \tilde{\mathbf{g}}_{n,t}\right\| _2^2$, $\forall n$. Set $ k_c\!=\!50 $, $ k\!=\!20 $.
	\item \textbf{Baseline 4}~\cite{scheduling2}: At each round, each device determines whether to be selected or not locally according to
\begin{equation*}
	P_n^t\!=\!\begin{cases}
	\label{equ:baseline3}
\left(\dfrac{\left| h_{n, t} \right|}{c\!+\!\left| h_{n,t} \right|^2} \right) ^2\!,&\left\| \tilde{\mathbf{g}}_{n,t}\right\| _2^2 \dfrac{\left| h_{n,t} \right|^2}{\left| h_{n,t} \right|^2\!+\!c}\! \geq\! c P_{on}, \\
	0,&else,
	\end{cases}
\end{equation*}
where $c\!=\!1$ indicates the energy cost in a certain scenario, and $P_{on}\!=\!4$ is the activation power for each device~\cite{scheduling2}.
\end{enumerate}


\vspace{-2mm}
\subsection{Influence of Parameter Settings on Training Efficiency}
\vspace{-0.5mm}

In this subsection, we analyze the influences of parameters $ \alpha $, $\lambda_E$/$\lambda_V$, and $ \gamma_\text{thr} $ on training performance by applying the proposed device scheduling mechanism, which can help guide the parameter settings in distinct situations.

\subsubsection{Influence of $ \alpha $}

We first analyze the influence of Lyapunov factor $ \alpha $ on the quantity of selected devices and the average energy consumption of each device over total communication rounds $T$,
the simulation results of which are shown in Figs.~\ref{fig:alpha_num} and~\ref{fig:alpha_energy}, respectively.
As illustrated in Fig.~\ref{fig:alpha_num}, the quantity of selected devices increases with increasing $ \alpha $, which results from that the edge server tends to select more devices to reduce the impact of noise, according to the optimization objective of device scheduling problem defined as (\ref{equ:opt:FCP:obj4}), (\ref{equ:U}) and $I_{n,t}\!=\!\lambda_EE_{n,t}-\lambda_VV_{n,t} $.
However, the cost of such preference is the increasing transmit energy consumption of all devices in the FEEL system, as shown in Fig.~\ref{fig:alpha_energy}.
To be specific, according to (\ref{equ:opt:FCP:obj4}), increasing $\alpha$ means the server is more inclined to increase the quantity of selected devices defined by $U_t$ than the quality of devices defined by $I_{n,t}$.
In other words, as $\alpha$ decreases, the server tends to select those devices with ``good'' qualities with large values of $I_{n,t}$, to resist the noise with a limited quantity.
Results in Fig.~\ref{fig:alpha} also indicate that the quantity of selected devices increases with $\sigma_0^2$ decreasing, which reflects the fact that when the channel condition is satisfied, the server tends to select more devices to participate FEEL to reduce the impact of noise and improve the accuracy and convergence of global model training.
Specifically, when $\sigma_0^2$ decreases, the transmit energy will be reduced resulting from the channel inversion based power control policy, as shown in Fig.~\ref{fig:alpha_energy}. Then the energy constraint formulated in (\ref{equ:opt:FCP:con2}) can be satisfied even with an increasing quantity of selected devices.
On the contrary, when the channel condition is not sufficiently satisfied, the server tends to select less devices with higher transmit energy to reduce the impact of channel distortion and ensure a good training performance.

Then we record the changing test accuracies from communication round $t=1$ to $t=T$ in Fig.~\ref{fig:alpha_acc}, under different $ \alpha $ settings and channel conditions to validate the above conclusions.
As presented in Fig.~\ref{fig:alpha_acc}, the test accuracies of global model under the three channel conditions can all convergence with relatively good performances,
which also reflects the adaptability and robustness of the proposed algorithm against different channel conditions. 
In addition, we observe that the test accuracies increase with increasing $ \alpha $, which is benefited from the larger quantity of selected devices and greater transmit energy consumption, as shown in Figs.~\ref{fig:alpha_num} and~\ref{fig:alpha_energy}.

\subsubsection{Influence of $\lambda_E$ and $\lambda_V$}
\label{lambda}

We further explore that how the distributed learning performance of proposed mechanism changes with $\lambda_E$ and $\lambda_V$.
Set $\lambda_E\!=\!0.3$, $0.4$, $0.5$, $0.6$ and $0.7$, and $\lambda_V\!=\!1\!-\!\lambda_E$.
According to the definition of $ I_{n,t}\!=\!\lambda_VV_{n,t}\!-\!\lambda_EE_{n,t} $, a larger value of $ \lambda_E $ means the optimization targets puts more emphasis on the energy consumption than the local updates and channel conditions, and then fewer devices will be selected.
However, the small quantity of selected devices will degrade the training accuracy.
As shown in Fig.~\ref{fig:lambda}, the quantity of selected devices, average energy consumption and test accuracy all decrease with increasing $\lambda_E$, which validates the analysis above.
Based on these results, we note that a proper setting of $ \lambda_E $ and $ \lambda_V $ can achieve a balance between the energy consumption and training accuracy, which can help improve the energy-efficiency of the FEEL system.

\subsubsection{Influence of $\gamma_{\text{thr}}$}

\begin{figure*}[!t]
	\begin{center}
		\subfigure[Quantity of selected device]{\includegraphics[width=0.31\textwidth]{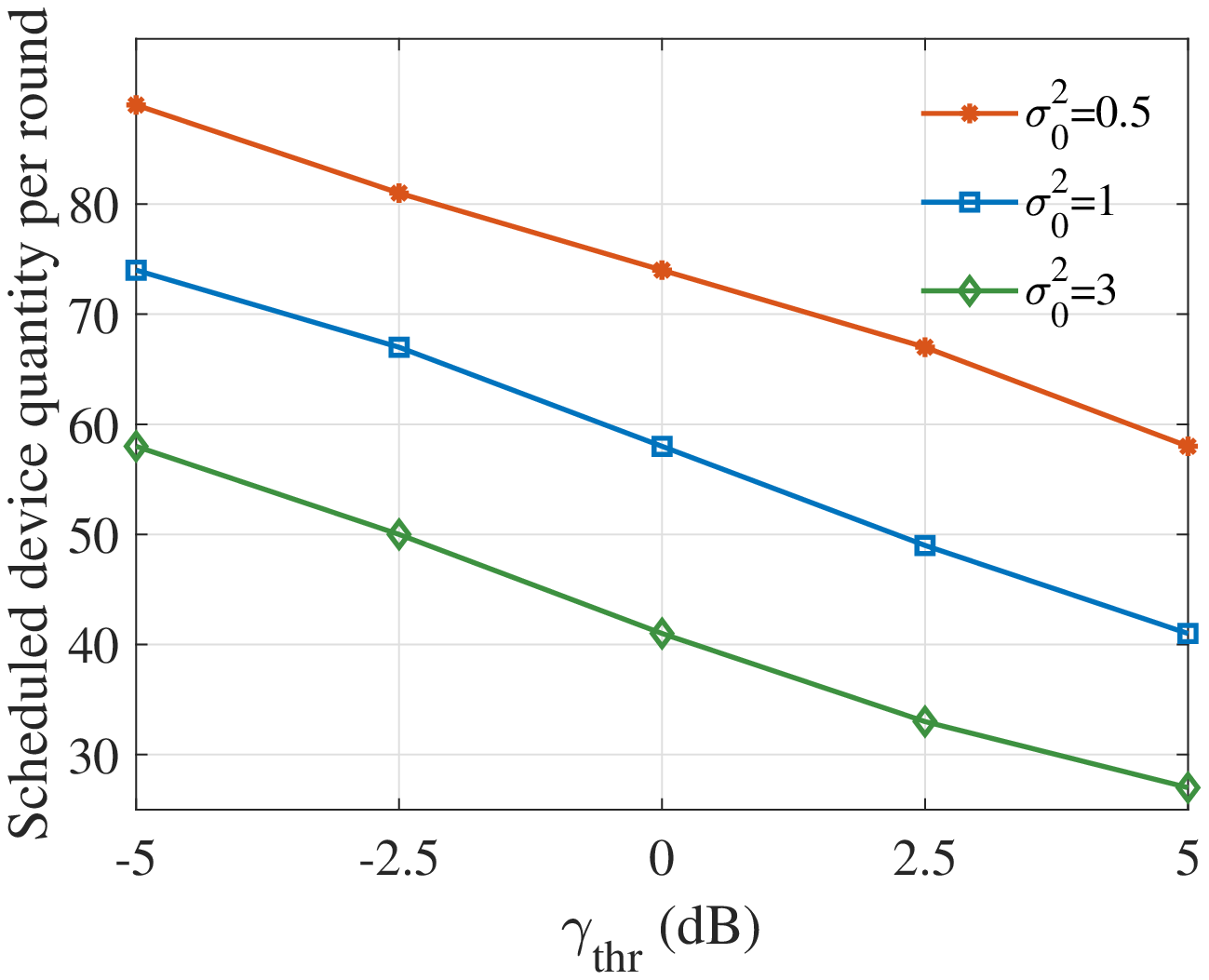}
			\label{fig:gamma_thr_quantity}}
		\subfigure[Average energy consumption ($ \sigma_0^2 = 1 $)]{\includegraphics[width=0.31\textwidth]{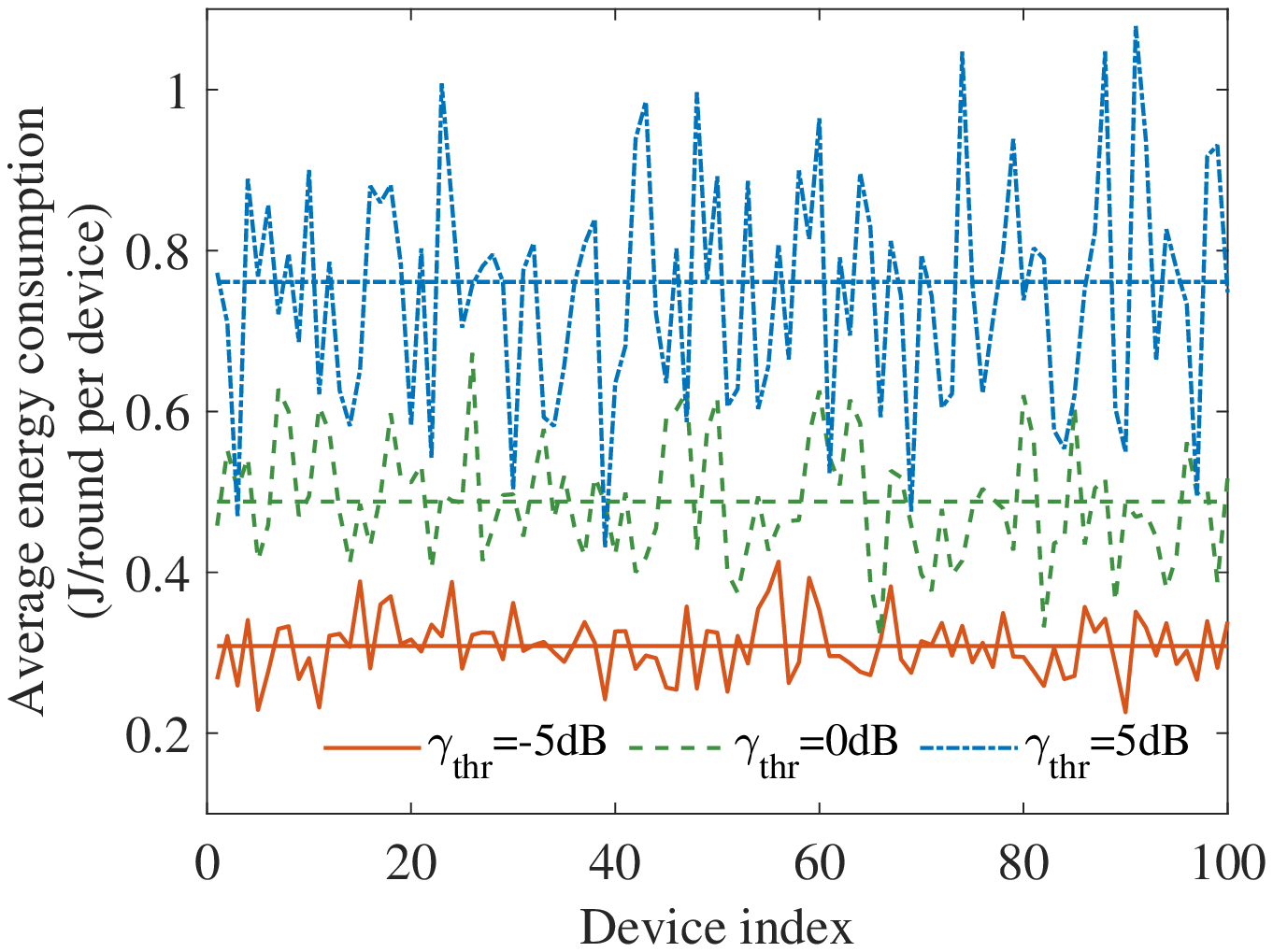}
			\label{fig:gamma_thr_energy}}
		\subfigure[Test accuracy ($ \sigma_0^2 = 1 $)]{\includegraphics[width=0.31\textwidth]{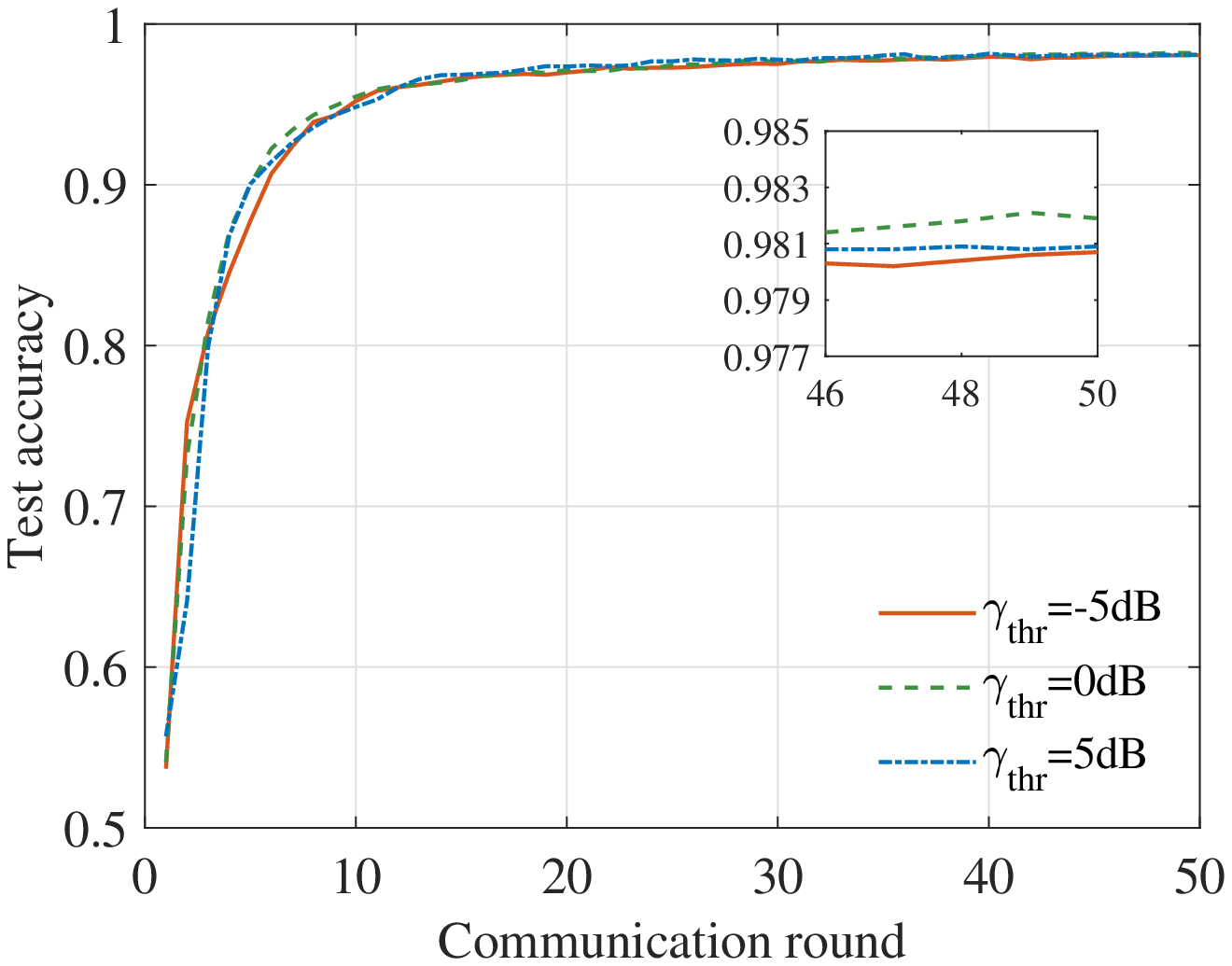}
			\label{fig:gamma_thr_acc}}
		\caption{Quantity of selected devices, average energy consumption and test accuracy versus $ \gamma_{thr} $ and $\sigma_0^2$ on MNIST dataset.}\vspace{-2mm}
		\label{fig:gamma_thr}
	\end{center}
\end{figure*}

\begin{figure*}[!t]
	\begin{center}
		\subfigure[$ \sigma_0^2 =0.5$]{\includegraphics[width=0.31\textwidth]{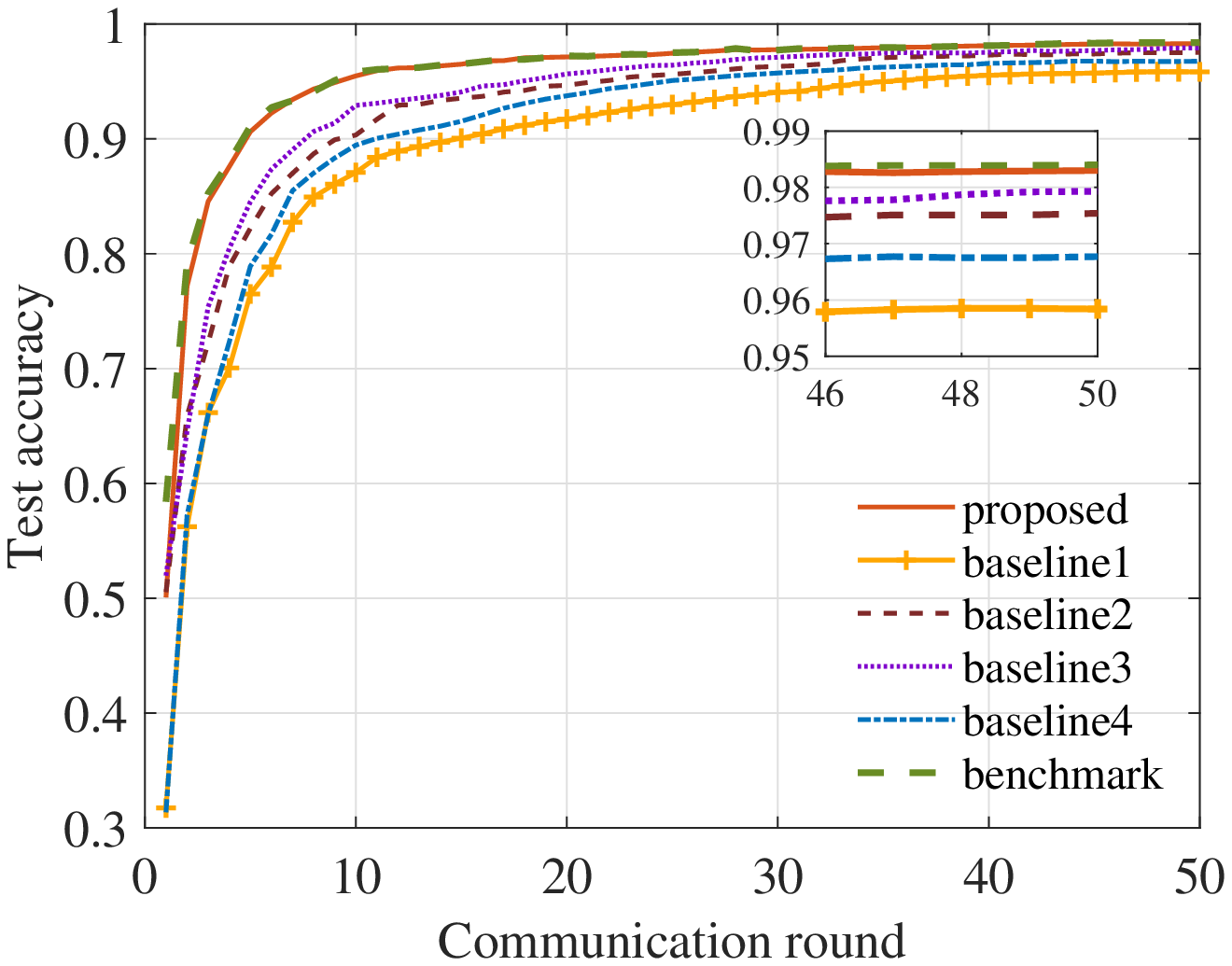}
			\label{fig:acc05}}
		\subfigure[$ \sigma_0^2 =1$]{\includegraphics[width=0.31\textwidth]{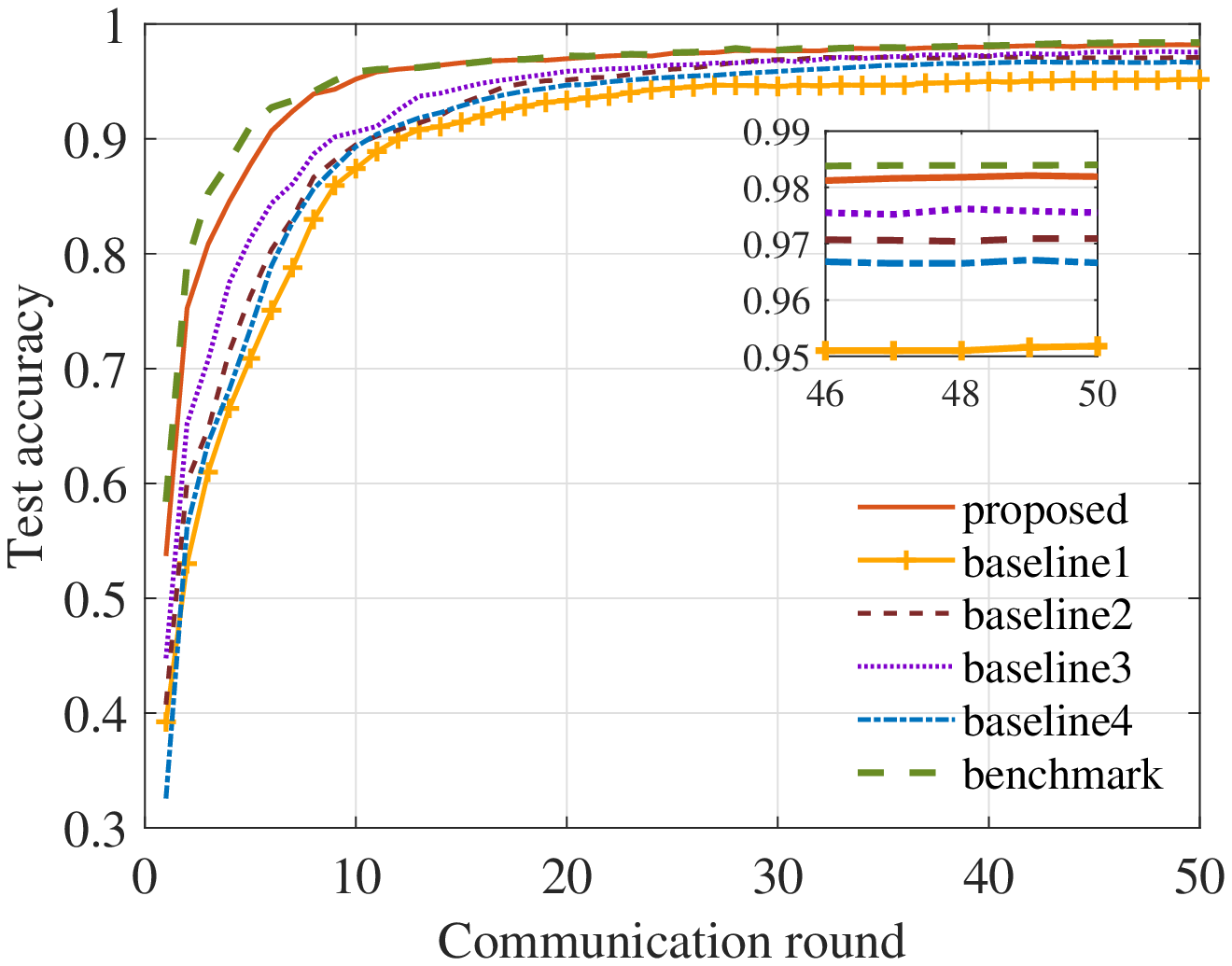}
			\label{fig:acc1}}
		\subfigure[$ \sigma_0^2 =3$]{\includegraphics[width=0.31\textwidth]{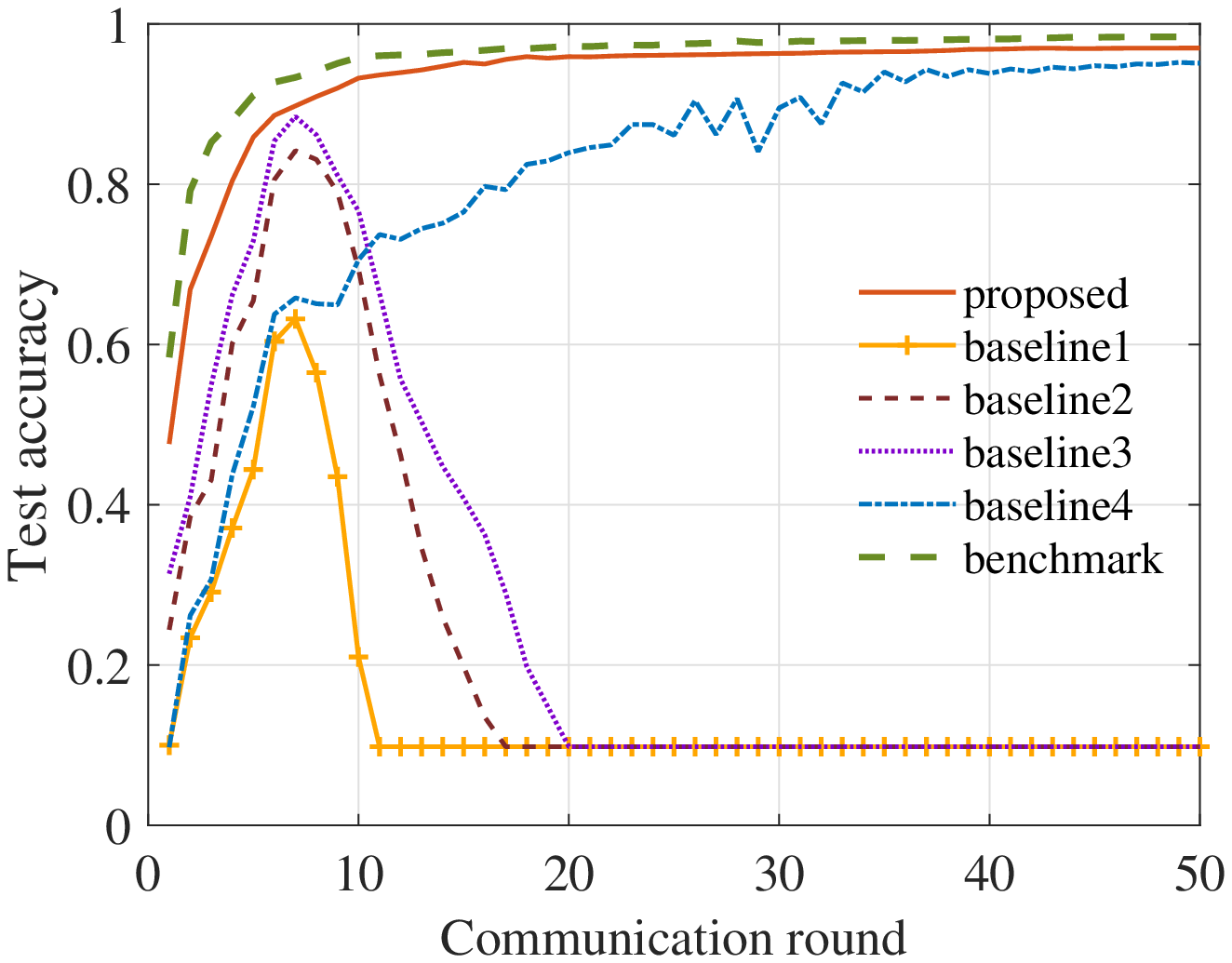}
			\label{fig:acc3}}
		\caption{Convergence of test accuracy under different channel condition on MNIST dataset.}\vspace{-2mm}
		\label{fig:acc}
	\end{center}
\end{figure*}

Then we analyze the influence of the received SNR threshold $ \gamma_{thr} $ on the quantity of selected devices, average energy consumption and test accuracy when applying the proposed dynamic scheduling mechanism. 
$ \gamma_{thr} $ selects values from $\left\{-5, -2.5, 0, 2.5, 5\right\}$ (dB), and results are shown in Fig.~\ref{fig:gamma_thr}.
In Fig.~\ref{fig:gamma_thr_quantity}, the quantity of selected devices decreases with growing $ \gamma_{thr} $ and deteriorating channel condition denoted by increasing $\sigma_0^2$, which reflects the fact that with growing $ \gamma_{thr} $, less devices will satisfy the higher standard of received SNR and cannot be selected by the server.
In addition, (\ref{equ:sigma}) requires larger transmit power with a higher $ \gamma_{thr} $, which leads to the higher energy consumption, as validated by Fig.~\ref{fig:gamma_thr_energy}.
Moreover, results in Fig.~\ref{fig:gamma_thr_acc} reflects that larger $ \gamma_{thr} $ may not promise a higher accuracy of model training.
To be specific, a small quantity of devices will be selected when $ \gamma_{thr} $ is large, which reduces the effectiveness of model training. On the other hand,
the influence of cumulative noise can be degraded by setting a small value of $ \gamma_{thr} $, especially when the channel condition is not sufficiently satisfied.

\subsection{Training Performance Improvements}
\label{simu1}

\begin{figure*}[!t]
	\begin{center}
		\subfigure[$ \sigma_0^2 =0.5$]{\includegraphics[width=0.31\textwidth]{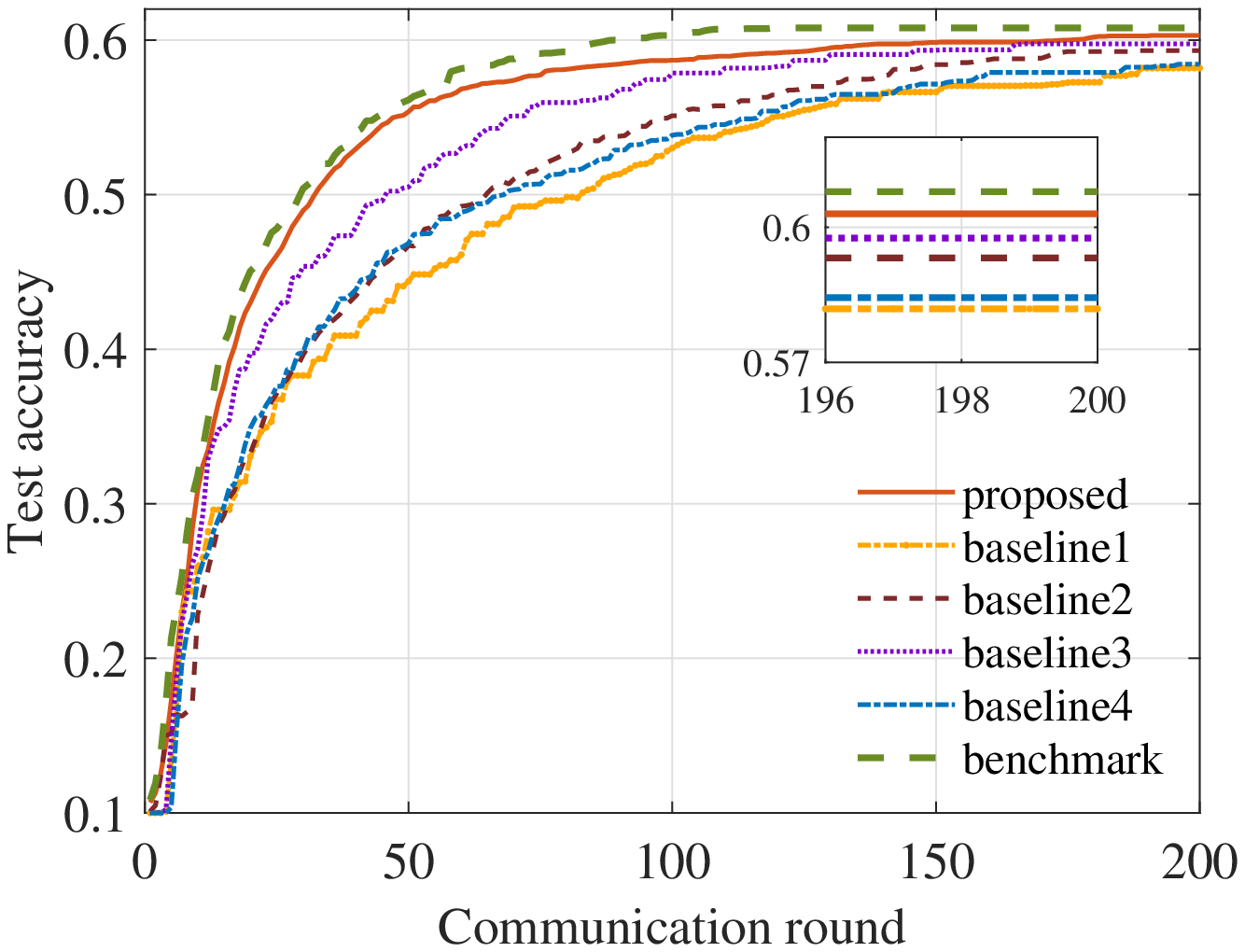}
			\label{fig:acc05cifar}}
		\subfigure[$ \sigma_0^2 =1$]{\includegraphics[width=0.31\textwidth]{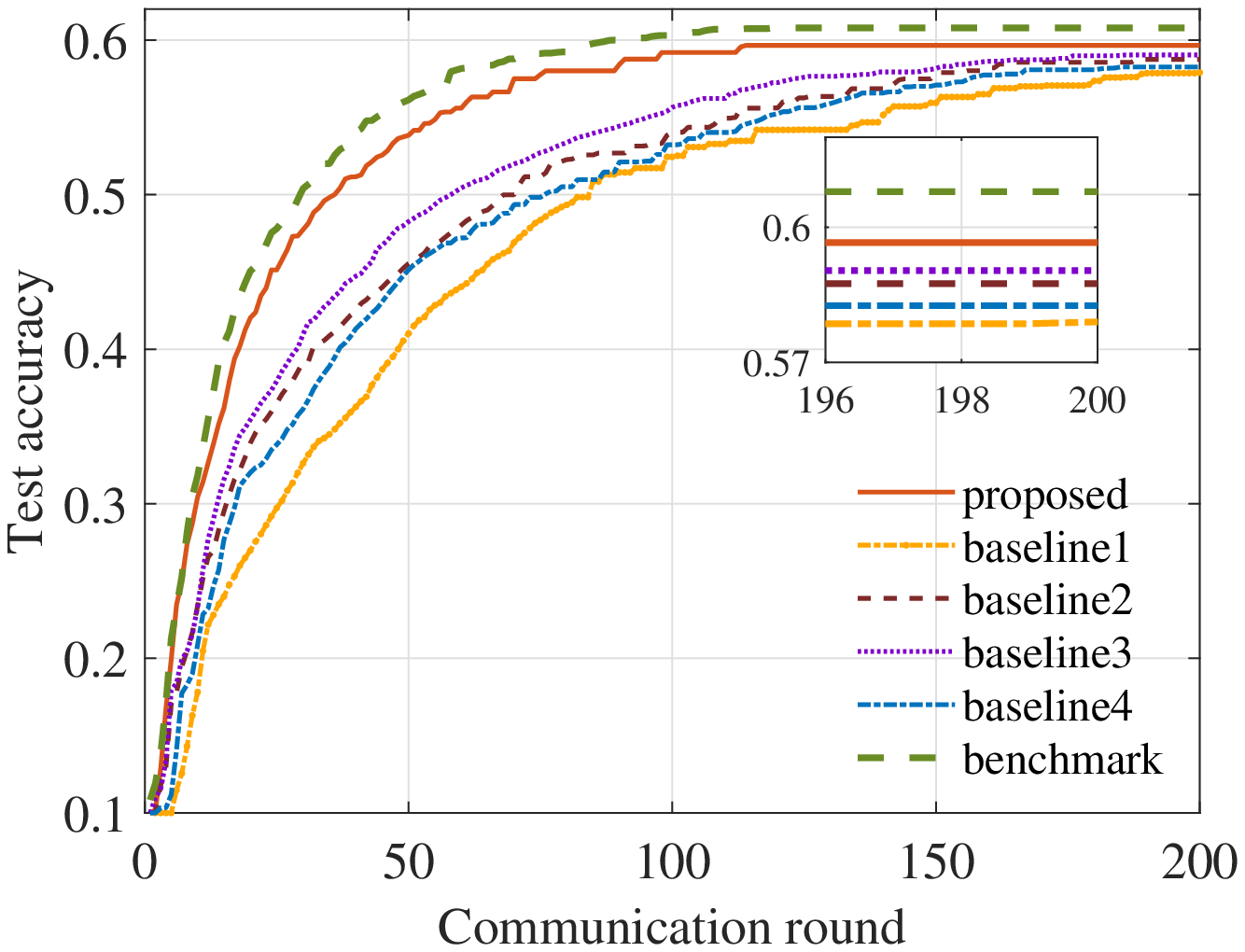}
			\label{fig:acc1cifar}}
		\subfigure[$ \sigma_0^2 =3$]{\includegraphics[width=0.31\textwidth]{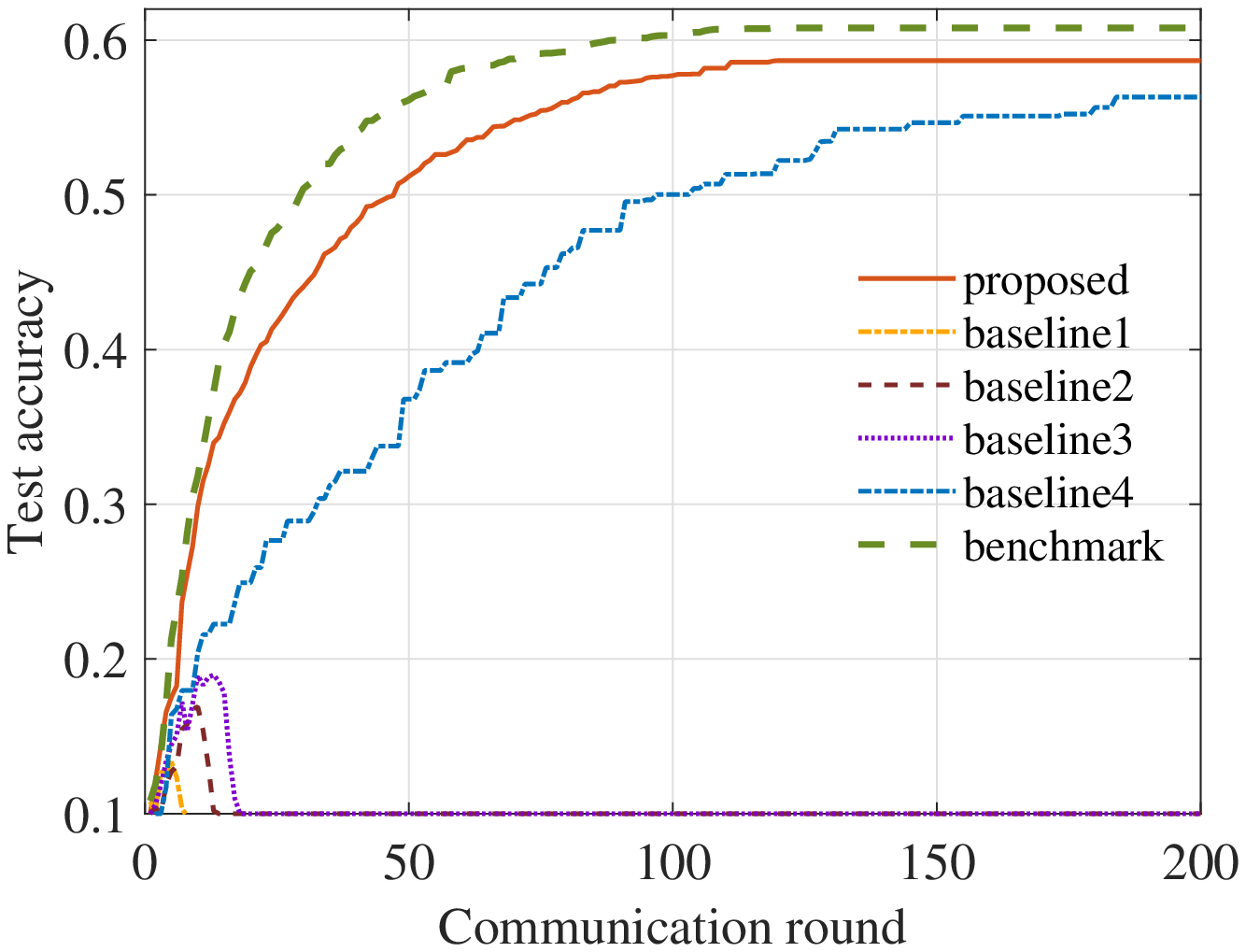}
			\label{fig:acc3cifar}}
		\caption{Convergence of test accuracy under different channel conditions on CIFAR-10 dataset.}\vspace{-2mm}
		\label{fig:acccifar}
	\end{center}
\end{figure*}


We further compare the test accuracy of the proposed mechanism with baselines introduced in Section~\ref{settings} on both MNIST and CIFAR-10 datasets.
Specifically, the CIFAR-10 dataset is complex and the training performance is more vulnerable to channel distortion.
Therefore, the training performance in Fig.~\ref{fig:acc} is better than that in Fig.~\ref{fig:acccifar}.
In addition, results in Figs.~\ref{fig:acc} and~\ref{fig:acccifar} represent that the proposed mechanism achieves the highest test accuracy with the fastest convergence rate on MNIST and CIFAR-10 datasets, which is benefited from the accumulated residual feedback, SNR-threshold-based power control and dynamic gradient-and-channel aware scheduling mechanism.
When the channel condition is satisfied, i.e., $ \sigma_0^2 =0.5$ and $ \sigma_0^2 =1$, 
the improvement of the proposed mechanism over baselines is marginal.
However, when $ \sigma_0^2 =3$, the training performance of baseline 1, baseline 2 and baseline 3 deteriorates rapidly,
while the proposed mechanism can ensure the convergence performance.
We observe that the random device selection scheme in baseline 1 has the worst performance as it does not consider any difference between devices, which results in inefficient energy consumption.
In addition, baseline 2 only adapts to the CSI, and baseline 3 considers CSI and DSI separately, both of which fail to consider the CSI and DSI jointly in the scheduling mechanism, and then results in low quality of selected devices.
Although baseline 4 exploits both the CSI and DSI jointly in its scheduling, the effect of channel fading is not well eliminated since its transmit power is fixed and unable to adapt to the changes of channel effectively.
Furthermore, the quantity of selected devices is independent of channel condition in all the baselines, which makes these baselines fail to adapt to different and changing channel noise.
In summary, the proposed device scheduling mechanism presents the best performance of test accuracy and robustness under different channel conditions compared with the baselines, and can achieve the close-to-optimal test accuracy indicated by the benchmark.

\begin{figure}[!t]
	\centering
	\includegraphics[width=0.469\textwidth]{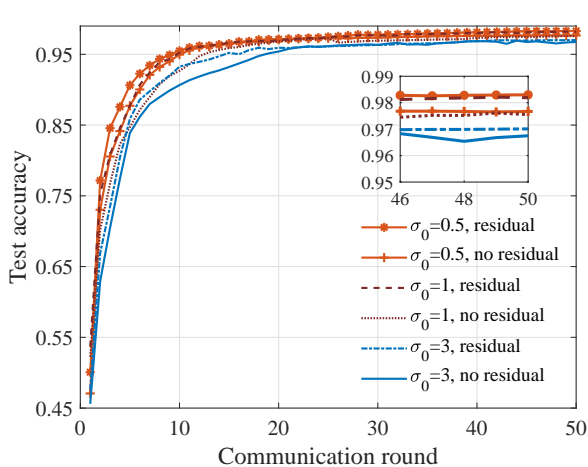}\vspace{-2mm}
	\caption{Convergence of test accuracy with/without residual feedback scheme under different channel conditions on MNIST dataset..}\vspace{-2mm}
	\label{fig:residual}
\end{figure}

Next, we continue to validate the benefit of the dynamic residual feedback scheme in Fig.~\ref{fig:residual}.
As shown in Fig.~\ref{fig:residual}, we can observe that the test accuracy and convergence rate with dynamic residual feedback scheme are both improved under the three channel conditions.
To be specific, there always exists a set of edge devices which can not participate the training process at each round due to the device selection.
Thus, the updates of unselected devices are not utilized sufficiently for global model updating, which would induce training bias and further affect the efficiency and accuracy of model training.
The dynamic residual feedback scheme can preserve the gradients of unselected devices and accumulate them for further gradient updating when the data and channel qualities are both satisfied.
Therefore, the dynamic residual feedback scheme enables the proposed mechanism to achieve higher test accuracy with faster convergence rate.


\vspace{-2mm}
\subsection{Proposed Mechanism in Non-i.i.d. Data Distribution}
\label{non_iid}

\begin{figure}[!t]
	\begin{center}
		\subfigure[Test accuracy on MNIST dataset.]{\includegraphics[width=0.469\textwidth]{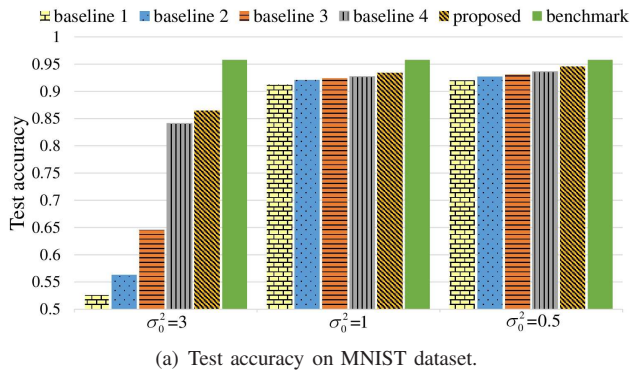}
			\label{fig:noniid_mnist}}
		\subfigure[Test accuracy on CIFAR-10 dataset.]{\includegraphics[width=0.469\textwidth]{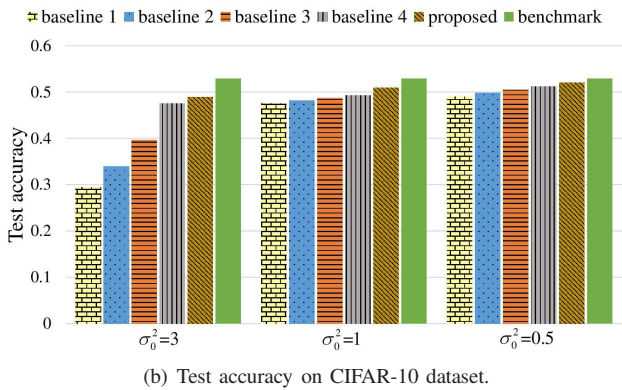}
			\label{fig:noniid_cifar}}
		\caption{Test accuracy in non-i.i.d. dataset under different channel conditions.}\vspace{-5mm}
		\label{fig:noniid}
	\end{center}
\end{figure}

Here we compare the performance of the proposed scheduling mechanism with baselines on non-i.i.d. datasets, as shown in Fig.~\ref{fig:noniid}.
For all evaluated mechanisms, the impact of channel noise on test accuracy is more aggressive in non-i.i.d. datasets than previous i.i.d datasets.
When $ \sigma_0^2 $ increases, our proposed mechanism still outperforms the baselines since the accumulated residual mechanism assists to use more important data in non-i.i.d. dataset, and the dynamic device scheduling with power control policy enables adaptiveness to different channel noise levels.
Furthermore, baseline 4 outperforms other baselines in terms of the test accuracy level, which also indicates that the consideration of both CSI and DSI plays an important role to improve the performance of device scheduling for the non-i.i.d. scenario.



\section{Conclusion}
\label{conculsion}
\vspace{-0.1mm}

In this paper, an AirComp enabled FEEL system has been established, which achieves communication-efficient distributed learning, and data aggregation can be achieved by exploiting the superposition property of multiple-access channels.
Then a gradient and channel aware dynamic device scheduling mechanism was designed to select qualified edge devices to participate the local data aggregation and global model updating,
considering the data update quality, channel condition, as well as energy consumption.
In this delicately designed mechanism, a power control scheme based on channel inversion was introduced to resist the channel fading and noise, and meet the threshold requirement of received SNR. In addition, an accumulated residual feedback based gradient transmission mechanism was investigated to improve the efficiency and accuracy of global model aggregation and updating.
Moreover, to solve the device scheduling problem through a distributed and online approach, a Lyapunov-drift-minimization-based algorithm was designed.
Furthermore, this work has characterized the convergence of the proposed device scheduling mechanism in the established FEEL system.

Simulations on MNIST and CIFAR-10 datasets have validated that the proposed mechanism improved the test accuracy and convergence rate, and was robust against different channel conditions for both i.i.d. and non-i.i.d. cases by comparing with different baselines.
Specifically, when the channel condition is not satisfied, a small quantity of contributive devices will be selected to upload updates with high transmit power to reduce the impact of channel distortion.
On the contrary, a large quantity of devices with low transmit power will be scheduled, which can achieve both efficient energy consumption and accurate model training.
As for future works, 
it is desirable to consider multi-frequency channels 
for a large-scale AirComp enabled FEEL system.
Furthermore, privacy protection and security issues are also worth investigating 
and more insecure circumstances can be considered.

\appendices
\vspace{-0.5mm}
\section{Proof of Lemma 1}
\label{proof:lemma1}
\begin{proof}
	The average residual can be separated as two parts:
	\begin{eqnarray}
	\begin{aligned}
	E\left\lbrack {\frac{1}{N}{\sum_{n = 1}^{N}\mathbf{r}_{n,t+1}}} \right\rbrack \leq E\left\lbrack {\frac{1}{N}\mathbf{r}_{t + 1}^{A}} \right\rbrack + E\left\lbrack {\frac{1}{N}\mathbf{r}_{t + 1}^{I}} \right\rbrack,
	\end{aligned}
	\end{eqnarray}
where $ \mathbf{r}_{t + 1}^{A} $ indicates the accumulated residual of selected devices and $ \mathbf{r}_{t + 1}^{I} $ indicates the accumulated residual of unselected devices. Therefore, we can obtain that
\begin{eqnarray*}
	\begin{aligned}
	E\left\lbrack \frac{1}{N}\mathbf{r}_{t + 1}^{A} \right\rbrack = 0,
	\end{aligned}
\end{eqnarray*}
\vspace{-1mm}
\begin{eqnarray*}
	\begin{aligned}
	&~~E\left\lbrack {\frac{1}{N}\mathbf{r}_{t + 1}^{I}} \right\rbrack= \frac{1}{N}E\left\lbrack {\sum\nolimits_{n \notin \mathbf{S}_{t}}\left| \left| \overset{\sim}{\mathbf{g}}_{n,t} \right| \right|_{2}^{2}} |  a_{n,t}=0 \right\rbrack\\	
	&\leq \frac{1}{N}\!\sum_{n \!=\! 1}^{N}\!E\left[ \left| \left| \tilde{\mathbf{g}}_{n,t} \right| \right|_{2}^{2} \right]\! Pr_t\!\leq \!\frac{1}{N}\!{\sum_{n = 1}^{N}}E\!\left[ \left\|  \mathbf{g}_{n,t} \right\|_{2}^{2} \!+\! \left\|  \mathbf{r}_{n,t} \right\|_{2}^{2} \right]\! Pr_t\\
	&\leq Pr_t\left( G^{2}+\left\|  \mathbf{g}_{t} \right\| _{2}^{2}\right)  + \frac{Pr_t}{N}{\sum_{n = 1}^{N}}E\left\lbrack \left| \left| \mathbf{r}_{n,t} \right| \right|_{2}^{2} \right\rbrack\\
	&\leq \frac{Pr_t\left( G^{2}+\left\|  \mathbf{g}_{t} \right\| _{2}^{2}\right)}{1 - Pr_t}.
	\end{aligned}
\end{eqnarray*}
\vspace{-0.5mm}
%
where $ Pr_t $ denotes the average possibility of that the device is unselected in round $t$, which is
determined by the channel condition.
This completes the proof of Lemma~\ref{lemma1}.
\end{proof}

\vspace{-5mm}
\section{proof of Lemma 2}
\label{proof:lemma2}
\begin{proof}
\vspace{-1mm}
	
According to (\ref{quantity}) and (\ref{equ:residual}), we have
\vspace{-1mm}
	\begin{eqnarray}
	\begin{aligned}
	\mathbf{w}_t =  \mathbf{w}_{t-1}-\zeta_t\left( \bar{\mathbf{g}}_t+\mathbf{r}_t + \hat{\bm{\varepsilon}}_t \right) ,
	\end{aligned}
	\end{eqnarray}
\vspace{-1mm}
where 
$ \bar{\mathbf{g}}_t \!= \!\frac{1}{\left|\mathbf{S}_t \right| }\!\sum\nolimits_{n\in \mathbf{S}_t}\!\mathbf{g}_{n,t} $,
$\mathbf{r}_t\!=\!\frac{1}{\left|\mathbf{S}_t \right| }\!\sum\nolimits_{n\in \mathbf{S}_t}\!\mathbf{r}_{n,t} $,
and $\hat{\bm{\varepsilon}}_t\! = \!\Phi^{-1}\left( \frac{\left[\bm{\varepsilon}_t \right]}{\sigma_t\left| \mathbf{S}_t\right| }\right) $.
According to assumptions in Section~\ref{convergence_assumption}, we can obtain that
\begin{eqnarray*}
	\begin{aligned}
	f&\left( \mathbf{w}_{t} \right)\! -\! f\left( \mathbf{w}_{t - 1} \right) \!\leq \!\nabla{f\left( \mathbf{w}_{t - 1} \right)}^{T}\left( {\mathbf{w}_{t}\! -\! \mathbf{w}_{t - 1}} \right) \!+\! \frac{L}{2}\left\|  \mathbf{w}_{t} \!-\!\mathbf{w}_{t - 1} \right\| _{2}^{2},\\	
	&\leq - \zeta_{t}\mathbf{g}_{t}^{T}\left( \bar{\mathbf{g}}_{t} \!+\! \mathbf{r}_{t} \right) \!-\! \zeta_{t}\mathbf{g}_{t}^{T}\hat{\bm{\varepsilon}}_{t} \!+\!  \frac{L{\zeta_{t}^{2}}}{2}\left( \left| \left| {\bar{\mathbf{g}}_{t} \!+\! \mathbf{r}_{t}} \right| \right|_{2}^{2}\!+\!\left\|  {\bm{\varepsilon}}_{t} \right\| _{2}^{2}  \right) .
	\label{ref:1}
	\end{aligned}
\end{eqnarray*}
\vspace{-1mm}
Considering that 
$ E_{\left|\mathcal{D}_{n,t}
	\right|}\!\left[  \left\|  \bar{\mathbf{g}}_{t} \right\| _{2}^{2} \right]  \!\leq\! \frac{G^{2}}{\left| \mathbf{S}_{t} \right| \left|\mathcal{D}_{n,t}^m\right|} \!+ \!\left\|  \mathbf{g}_{t}\right\|_{2}^{2} $~\cite{scheduling1},
$E\left[\left\| \hat{ \bm{\varepsilon}}_t\right\| ^2  \right]\!\leq\!\frac{\sigma_0^2 G^2}{\sigma_t^2 \left| \mathbf{S}_t\right|^2}\!+\!m^2 $, where $m$ is the estimate of the
global gradient mean, 
we take the expectation at both sides and have
\begin{eqnarray*}
	\begin{aligned}
	&E\left[ f\left( \mathbf{w}_{t} \right) - f\left( \mathbf{w}_{t - 1} \right) \right]\leq - \zeta_{t}\mathbf{g}_{t}^{T} \left( \bar{\mathbf{g}}_{t} + \mathbf{r}_{t} \right)- \zeta_{t}\mathbf{g}_{t}^{T}\hat{\bm{\varepsilon}}_{t}\\
	&+ \frac{L{\zeta_{t}^{2}}}{2}\left| \left| {\bar{\mathbf{g}}_{t} + \mathbf{r}_{t}} \right| \right|_{2}^{2} + \frac{L{\zeta_{t}^{2}}}{2}\left(\frac{\sigma_0^2 \delta^2}{\sigma_t^2 \left| \mathbf{S}_t\right|^2}\!+\!m^2\right)\\
	&\leq - \zeta_{t}\left\| \mathbf{g}_{t}\right\|_2^2 - \zeta_{t}\mathbf{g}_{t}^T \mathbf{r}_{t}- \zeta_{t}\mathbf{g}_{t}^{T}\hat{\bm{\varepsilon}}_{t}+ \frac{L\zeta_{t}^{2}}{2}\left( \frac{G^{2}}{\left| \mathbf{S}_{t} \right| \left|\mathcal{D}_{n,t}^m\right|} \!+ \!\left\|  \mathbf{g}_{t}\right\|_{2}^{2}\right) \\
	&+ \frac{L{\zeta_{t}^{2}}}{2}\left\| \mathbf{r}_{t} \right\| _{2}^{2} + \frac{L{\zeta_{t}^{2}}}{2}\left(\frac{\sigma_0^2 \delta^2}{\sigma_t^2 \left| \mathbf{S}_t\right|^2}\!+\! m^2\right)\\
	&\leq \left( \frac{L\zeta_{t}^{2}}{2}- \zeta_{t}\right) \left\| \mathbf{g}_{t}\right\|_2^2 - \zeta_{t}\mathbf{g}_{t}^T \mathbf{r}_{t}- \zeta_{t}\mathbf{g}_{t}^{T}\hat{\bm{\varepsilon}}_{t}+ \frac{L\zeta_{t}^{2}}{2}\frac{G^{2}}{\left| \mathbf{S}_{t} \right| \left|\mathcal{D}_{n,t}^m\right|}\\
	&+ \frac{L{\zeta_{t}^{2}}}{2}\left\| \mathbf{r}_{t} \right\| _{2}^{2} + \frac{L{\zeta_{t}^{2}}}{2}\left(\frac{\sigma_0^2 \delta^2}{\sigma_t^2 \left| \mathbf{S}_t\right|^2}\!+\! m^2\right).
	\end{aligned}
\end{eqnarray*}
\vspace{-1mm}
By applying $ -a^Tb\leq\left( \left\| a\right\| _2^2+\left\| b\right\| _2^2\right) /2$, then we have
\begin{eqnarray*}
	\begin{aligned}
	&E\left[ f\left( \mathbf{w}_{t} \right) - f\left( \mathbf{w}_{t - 1} \right) \right]\leq  \frac{L\zeta_{t}^{2}}{2} \left\| \mathbf{g}_{t}\right\|_2^2 + \frac{\zeta_{t}}{2}\left\| \mathbf{r}_{t}\right\| _2^2+\frac{\zeta_t}{2} \left\| \hat{\bm{\varepsilon}}_{t}\right\| _2^2\\
	&+\frac{L\zeta_{t}^{2}m^2}{2}+\frac{L\zeta_{t}^{2}}{2}\left\| \mathbf{r}_{t} \right\| _{2}^{2}+\frac{L\zeta_{t}^{2}}{2}\left(\frac{\delta^2}{\gamma_{thr} \left| \mathbf{S}_{t} \right|^{2}}\! +\! \frac{G^2}{\left| \mathbf{S}_{t} \right|\left|\mathcal{D}_{n,t}^m\right|} \right)\\
	&\leq  \frac{L\zeta_{t}^{2}}{2} \left\| \mathbf{g}_{t}\right\|_2^2+\frac{L\zeta_{t}^{2}+\zeta_t}{2}m^2+\frac{L\zeta_{t}^{2}+\zeta_t}{2}\left\| \mathbf{r}_{t} \right\| _{2}^{2}\\
	&+\frac{\zeta_{t}}{2}\frac{\delta^2}{\gamma_{thr} \left| \mathbf{S}_{t} \right|^{2}}+\frac{L\zeta_{t}^{2}}{2}\left(\frac{\delta^2}{\gamma_{thr} \left| \mathbf{S}_{t} \right|^{2}}\! +\! \frac{G^2}{\left| \mathbf{S}_{t} \right|\left|\mathcal{D}_{n,t}^m\right|} \right)\\
	&\overset{(a)}{\leq}  \frac{L\zeta_{t}^{2}}{2} \left\| \mathbf{g}_{t}\right\|_2^2+\frac{L\zeta_{t}^{2}+\zeta_t}{2}m^2+\frac{L\zeta_{t}^{2}+\zeta_t}{2}\frac{Pr_t\left( G^{2}+\left\|  \mathbf{g}_{t} \right\| _{2}^{2}\right)}{1 - Pr_t}\\
	&+\frac{\zeta_{t}}{2}\left[ \frac{\left( L\zeta_t+1\right) \delta^2}{\gamma_{thr} \left| \mathbf{S}_{t} \right|^{2}}+\frac{L\zeta_{t}G^2}{\left| \mathbf{S}_{t} \right|\left|\mathcal{D}_{n,t}^m\right|}\right] \\
	\end{aligned}
\end{eqnarray*}
where inequality $\left( a \right)$ is obtained according to Lemma~\ref{lemma1}.
Without loss of generality, we have $ L\zeta_t\!+\!1 \!\approx\! L\zeta_t $. Therefore,
\begin{eqnarray*}
	\begin{aligned}	
	&E\left[ f\left( \mathbf{w}_{t} \right)\right] - E\left[ f\left( \mathbf{w}_{t - 1} \right) \right] \leq \frac{L\zeta_{t}^{2}}{2} \left\| \mathbf{g}_{t}\right\|_2^2+\frac{L\zeta_{t}^{2}+\zeta_t}{2}m^2\\
	&+\!\frac{L\zeta_{t}^{2}\!+\!\zeta_t}{2}\!\frac{Pr_t\!\left(\! G^{2}\!+\!\left\|  \mathbf{g}_{t} \right\| _{2}^{2}\!\right)}{1 \!-\! Pr_t}\!+\!\frac{L\zeta_{t}^2}{2}\left[ \!\frac{\delta^2}{\gamma_{thr} \left| \mathbf{S}_{t} \right|^{2}}\!+\!\frac{G^2}{\left| \mathbf{S}_{t} \right|\left|\mathcal{D}_{n,t}^m\right|}\!\right].
	\label{ref:g}
	\end{aligned}
\end{eqnarray*}
This completes the proof of Lemma~\ref{lemma2}.
\end{proof}

\vspace{-3mm}
\section{Proof of Theorem 1}
\label{proof:theorem1}
\begin{proof}
	
Define the Lyapunov function as $ L_t=f\left( \mathbf{w}_t\right) - f^* $, where $ f $ is the loss function and $ f^* $ is the optimal value of $f$. Then the one-round Lyapunov drift term is
\begin{eqnarray*}
	\begin{aligned}
	\Delta_{1} \left(t \right)  =\! E\left[ L_{t}-L_{t-1}\right]=\! E\left[ f\left( \mathbf{w}_t\right)\! -\! f^*\!-\!f\left( \mathbf{w}_{t-1}\right)\! +\! f^* \right].
	\end{aligned}
\end{eqnarray*}
Therefore, we can derive the T-round drift term as
\begin{eqnarray*}
\begin{aligned}
\mathrm{\Delta}_{T} \!=\! E\left[  L_{T+1} - L_{1} \right]\!=\!\sum_{t=1}^{T}\Delta_1\left(t \right)\!=\!E\left[f\left(\mathbf{w}_T \right)-f\left(\mathbf{w}_0 \right)  \right]  .
\end{aligned}
\end{eqnarray*}
Based on the third assumption in Section~\ref{convergence_assumption}, we have
\begin{eqnarray*}
	\begin{aligned}
	\left\| \mathbf{g}_t\right\| _2^2 \geq 2 \mu   \left(E\left[f\left(\mathbf{w}_{t-1} \right)\right] -f^* \right).
	\label{ref:g}
	\end{aligned}
\end{eqnarray*}
Combining with the one round convergence rate, we have that
\begin{eqnarray*}
	\begin{aligned}
		&E\left[ f\left( \mathbf{w}_{t} \right)\right] - E\left[ f\left( \mathbf{w}_{t - 1} \right) \right] \leq \frac{L\zeta_{t}^{2}}{2} \left\| \mathbf{g}_{t}\right\|_2^2+\frac{L\zeta_{t}^{2}+\zeta_t}{2}m^2\\
		&+\!\frac{L\zeta_{t}^{2}\!+\!\zeta_t}{2}\!\frac{Pr_t\!\left(\! G^{2}\!+\!\left\|  \mathbf{g}_{t} \right\| _{2}^{2}\!\right)}{1 \!-\! Pr_t}\!+\!\frac{L\zeta_{t}^2}{2}\left[ \!\frac{\delta^2}{\gamma_{thr} \left| \mathbf{S}_{t} \right|^{2}}\!+\!\frac{G^2}{\left| \mathbf{S}_{t} \right|\left|\mathcal{D}_{n,t}^m\right|}\!\right]\\
		&\leq \left[ \frac{L\zeta_{t}^{2}}{2}+ \frac{L\zeta_{t}^{2}Pr_t\!+\!\zeta_t Pr_t}{2\left(1-Pr_t \right) }\right] \left\|\mathbf{g}_{t}\right\|_2^2
		+\frac{L\zeta_{t}^{2}+\zeta_t}{2}m^2\\
		&+\!\frac{L\zeta_{t}^{2}\!+\!\zeta_t}{2}\!\frac{Pr_tG^{2}\!}{1 \!-\! Pr_t}\!+\!\frac{L\zeta_{t}^2}{2}\left[ \!\frac{\delta^2}{\gamma_{thr} \left| \mathbf{S}_{t} \right|^{2}}\!+\!\frac{G^2}{\left| \mathbf{S}_{t} \right|\left|\mathcal{D}_{n,t}^m\right|}\!\right]\\
		&\leq \left[ \frac{L\zeta_{t}^{2}}{2}\!+\! \frac{\left( L\zeta_{t}^{2}\!+\!\zeta_t\right) Pr_t}{2\left(1\!-\!Pr_t \right) }\right]\! 2\mu \left(\!E\left[ f\left(\mathbf{w}_{t-1} \right)\!\right] \!-\!f^*\right)
		\!+\!\frac{L\zeta_{t}^{2}\!+\!\zeta_t}{2}m^2\\
		&+\!\frac{L\zeta_{t}^{2}\!+\!\zeta_t}{2}\!\frac{Pr_tG^{2}\!}{1 \!-\! Pr_t}\!+\!\frac{L\zeta_{t}^2}{2}\left[ \!\frac{\delta^2}{\gamma_{thr} \left| \mathbf{S}_{t} \right|^{2}}\!+\!\frac{G^2}{\left| \mathbf{S}_{t} \right|\left|\mathcal{D}_{n,t}^m\right|}\!\right].
	\end{aligned}
\end{eqnarray*}
Let $ B_t\!\triangleq\! \frac{L\zeta_{t}^{2}\!+\!\zeta_t}{2}m^2
+\!\frac{L\zeta_{t}^{2}\!+\!\zeta_t}{2}\!\frac{Pr_tG^{2}\!}{1 \!-\! Pr_t}\!+\!\frac{L\zeta_{t}^2}{2}\left[ \!\frac{\delta^2}{\gamma_{thr} \left| \mathbf{S}_{t} \right|^{2}}\!+\!\frac{G^2}{\left| \mathbf{S}_{t} \right|\left|\mathcal{D}_{n,t}^m\right|}\!\right]$, then we can derive that
\begin{eqnarray*}
	\begin{aligned}
		&E\left[ f\left( \mathbf{w}_{t} \right)\right] - E\left[ f\left( \mathbf{w}_{t - 1} \right) \right] \\
		&\leq \left[ \frac{L\zeta_{t}^{2}}{2}\!+\! \frac{\left( L\zeta_{t}^{2}\!+\!\zeta_t\right) Pr_t}{2\left(1\!-\!Pr_t \right) }\right]\! 2\mu \left(E\left[f\left(\mathbf{w}_{t-1} \right)\right] \!-\!f^*  \right)
		\!+\!B_t\\
		&E\left[ f\left( \mathbf{w}_t\right)\right]  -f^* \\
		&\leq \left[1 \!+\!2\mu\left(\frac{L\zeta_{t}^{2}}{2}\!+\! \frac{\left( L\zeta_{t}^{2}\!+\!\zeta_t\right) Pr_t}{2\left(1\!-\!Pr_t \right) } \right) 
		\right] \left( E\left[ f\left( \mathbf{w}_{t-1}\right)\right] -f^*\right)  \!+\!B_t.
	\end{aligned}
\end{eqnarray*}
Then let $ C_t\!\triangleq\! 1 \!+\!2\mu\left(\frac{L\zeta_{t}^{2}}{2}\!+\! \frac{\left( L\zeta_{t}^{2}\!+\!\zeta_t\right) Pr_t}{2\left(1\!-\!Pr_t \right) } \right)$, 
we can obtain
\begin{eqnarray*}
	\begin{aligned}
		&E\left[ f\left( \mathbf{w}_T\right)\right] -f^*\\
		&\leq C_T\left( E\left[ f\left( \mathbf{w}_{T-1}\right)\right] -f^*\right) +B_T\\
		&\leq C_T \left[ C_{T-1}\left(   E\left[f\left( \mathbf{w}_{T-2}\right)  \right] \!-\!f^*\right)  +B_{T-1} \right] \!+\!B_T\\
		& \dots \\
		&\leq \prod_{i=1}^{T} C_i \left[ E\left[ f\left(\mathbf{w}_0 \right)\right]\!-\!f^*\right]  \!+\!\sum_{i=1}^{T-1}B_i\prod_{j=i+1}^{t}C_j \!+\! B_T.
	\end{aligned}
\end{eqnarray*}
Therefore, we can derive the T-round drift term as
\begin{eqnarray*}
	\begin{aligned}
	&\mathrm{\Delta}_{T} = E\left[  f\left(\mathbf{w}_T \right)  -f\left(\mathbf{w}_0 \right)\right] \\
	&\leq \left( \prod_{i=1}^{T} C_i-1\right)  \left[ E\left[ f\left(\mathbf{w}_0 \right)\right]\!-\!f^*\right]  \!+\!\sum_{i=1}^{T-1}B_i\prod_{j=i+1}^{t}C_j \!+\! B_T.
	\end{aligned}
\end{eqnarray*}
This completes the proof of Theorem~\ref{theorem1}.
\end{proof}

\bibliographystyle{IEEEtran}
\bibliography{ref}

\begin{IEEEbiography}[{\includegraphics[width=1in,height=1.25in,clip,keepaspectratio]{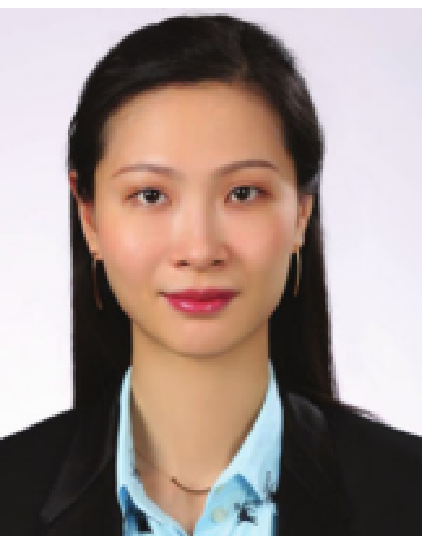}}]{Jun Du} (S'16-M'18-SM'21) received her B.S. in information and communication engineering from Beijing Institute of Technology, in 2009, and her M.S. and Ph.D. in information and communication engineering from Tsinghua University, Beijing, in 2014 and 2018, respectively. From Oct. 2016 - Sept. 2017, Dr. Du was a sponsored researcher, and she visited Imperial College London. Currently she is an assistant professor in the Department of Electrical Engineering, Tsinghua University. Her research interests are mainly in communications, networking, resource allocation and system security problems of heterogeneous networks and space-based information networks. Dr. Du is the recipient of the Best Student Paper Award from IEEE GlobalSIP in 2015, the Best Paper Award from IEEE ICC 2019, and the Best Paper Award from IWCMC in 2020.
\end{IEEEbiography}

\begin{IEEEbiography}[{\includegraphics[width=1in,height=1.25in,clip,keepaspectratio]{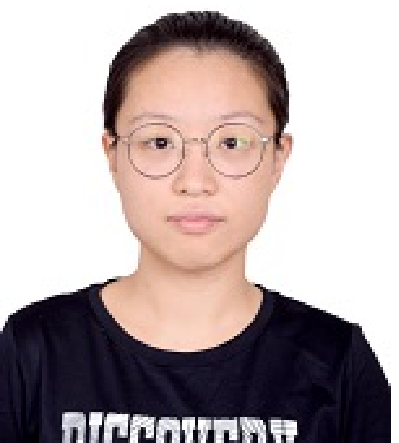}}]{Bingqing Jiang} (S’21) received the B.S. degree from the School of Information and Communication Engineering, Beijing University of Posts and Telecommunications, Beijing, China, in 2022. She is currently pursuing the M.S. degree in electronics and communication engineering from Tsinghua University, Beijing, China. Her research interests include machine learning and wireless communications.
\end{IEEEbiography}

\begin{IEEEbiography}[{\includegraphics[width=1in,height=1.25in,clip,keepaspectratio]{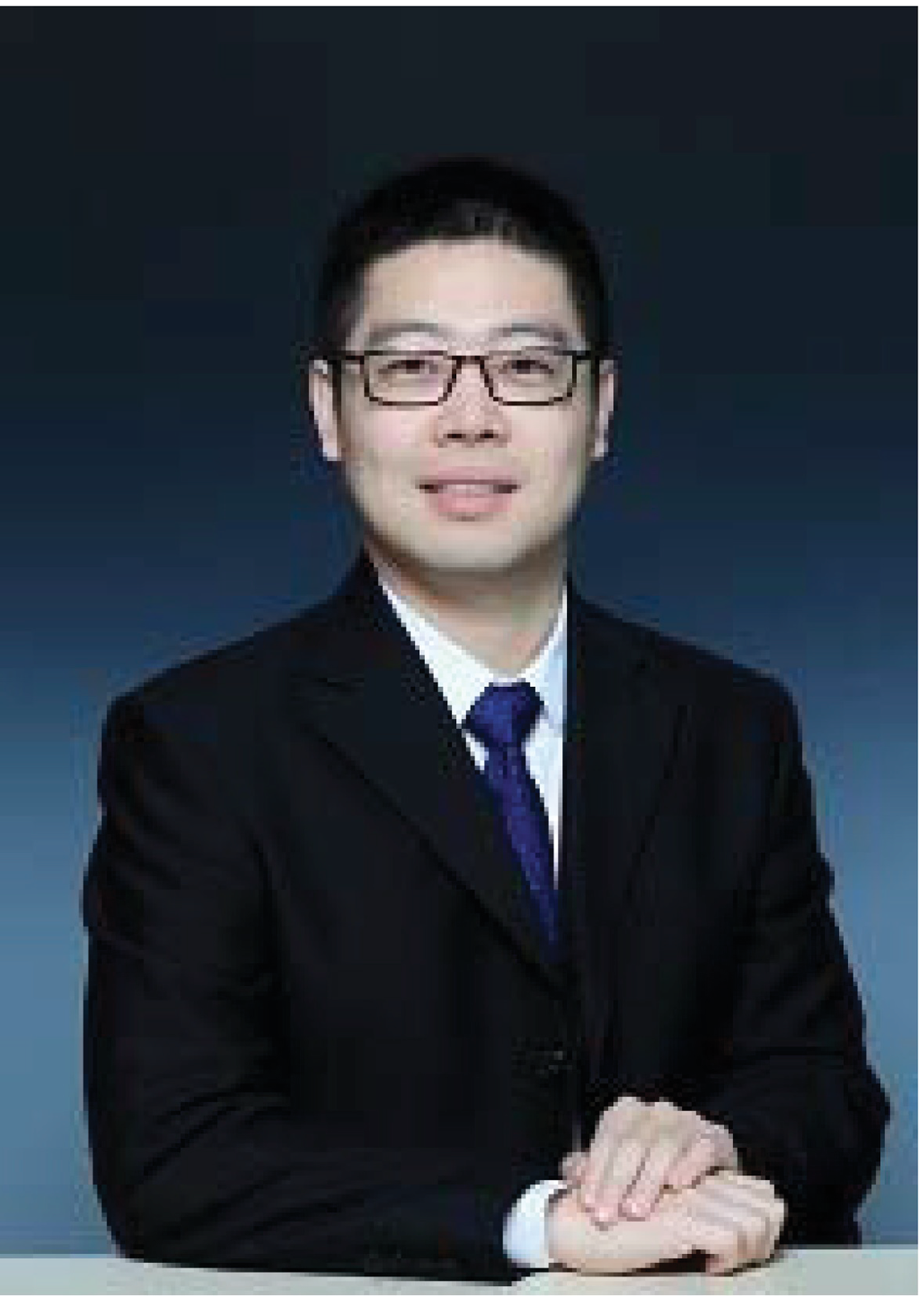}}]{Chunxiao Jiang} (S'09-M'13-SM'15) is an associate professor in School of Information Science and Technology, Tsinghua University. He received the B.S. degree in information engineering from Beihang University, Beijing in 2008 and the Ph.D. degree in electronic engineering from Tsinghua University, Beijing in 2013, both with the highest honors. From 2011 to 2012 (as a Joint Ph.D) and 2013 to 2016 (as a Postdoc), he was in the Department of Electrical and Computer Engineering at University of Maryland College Park under the supervision of Prof. K. J. Ray Liu. His research interests include application of game theory, optimization, and statistical theories to communication, networking, and resource allocation problems, in particular space networks and heterogeneous networks. Dr. Jiang has served as an Editor of IEEE Transactions on Communications, IEEE Internet of Things Journal, IEEE Wireless Communications, IEEE Transactions on Network Science and Engineering, IEEE Network, IEEE Communications Letters, and a Guest Editor of IEEE Communications Magazine, IEEE Transactions on Network Science and Engineering and IEEE Transactions on Cognitive Communications and Networking. He has also served as a member of the technical program committee as well as the Symposium Chair for a number of international conferences. Dr. Jiang is the recipient of the Best Paper Award from IEEE GLOBECOM in 2013, IEEE Communications Society Young Author Best Paper Award in 2017, the Best Paper Award from ICC 2019, IEEE VTS Early Career Award 2020, IEEE ComSoc Asia-Pacific Best Young Researcher Award 2020, IEEE VTS Distinguished Lecturer 2021, and IEEE ComSoc Best Young Professional Award in Academia 2021. He received the Chinese National Second Prize in Technical Inventions Award in 2018 and Natural Science Foundation of China Excellent Young Scientists Fund Award in 2019. He is a Senior Member of IEEE and a Fellow of IET.
\end{IEEEbiography}

\begin{IEEEbiography}[{\includegraphics[width=1in,height=1.25in,clip,keepaspectratio]{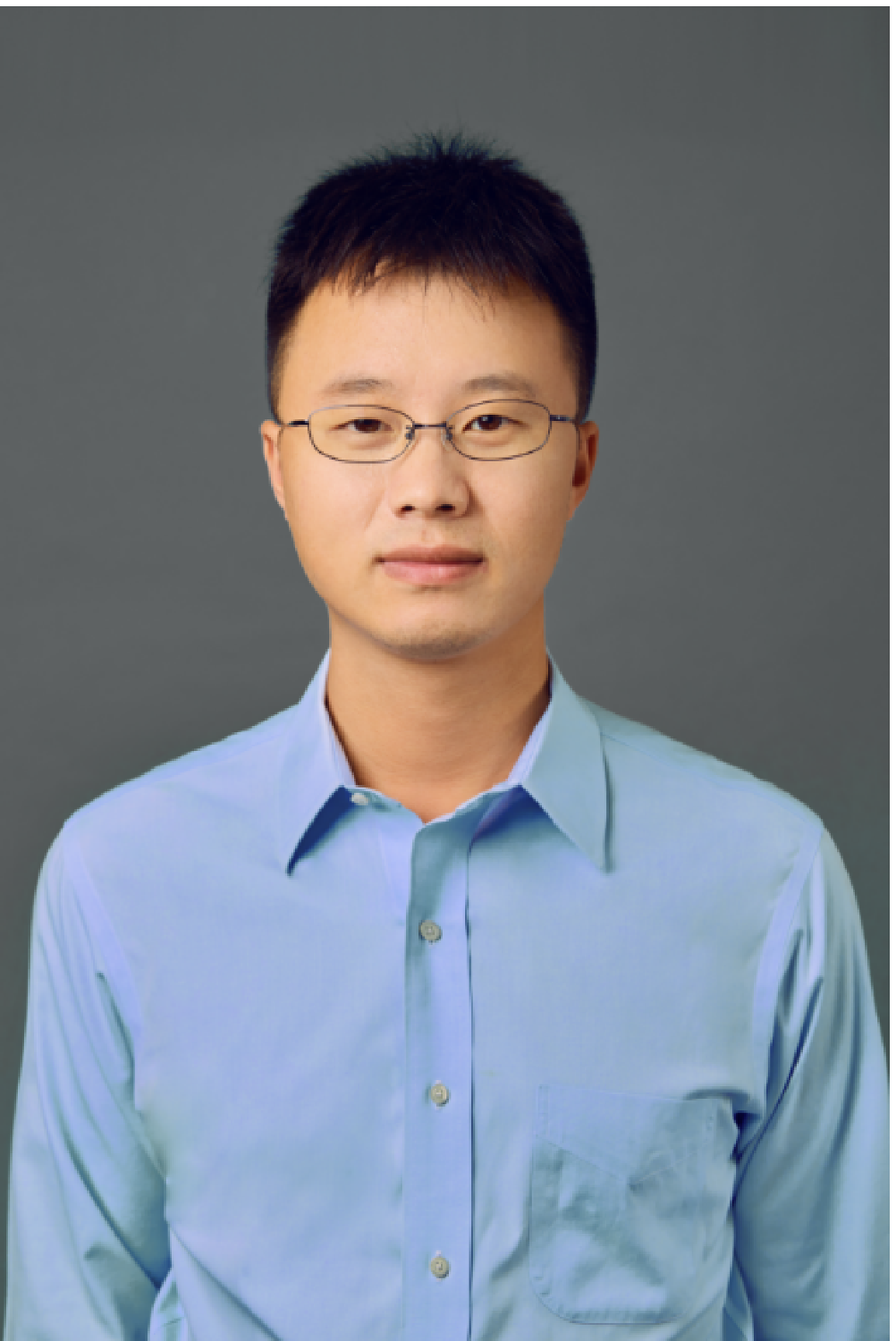}}]{Yuanming Shi} (S’13-M’15-SM’20) received the B.S. degree in electronic engineering from Tsinghua University, Beijing, China, in 2011. He received the Ph.D. degree in electronic and computer engineering from The Hong Kong University of Science and Technology (HKUST), in 2015. Since September 2015, he has been with the School of Information Science and Technology in ShanghaiTech University, where he is currently a tenured Associate Professor. He visited University of California, Berkeley, CA, USA, from October 2016 to February 2017. His research areas include optimization, machine learning, wireless communications, and their applications to 6G, IoT, and edge AI. He was a recipient of the 2016 IEEE Marconi Prize Paper Award in Wireless Communications, the 2016 Young Author Best Paper Award by the IEEE Signal Processing Society, and the 2021 IEEE ComSoc Asia-Pacific Outstanding Young Researcher Award. He is also an editor of IEEE Transactions on Wireless Communications, IEEE Journal on Selected Areas in Communications, and Journal of Communications and Information Networks.
	
\end{IEEEbiography}

\begin{IEEEbiography}[{\includegraphics[width=1in,height=1.25in,clip,keepaspectratio]{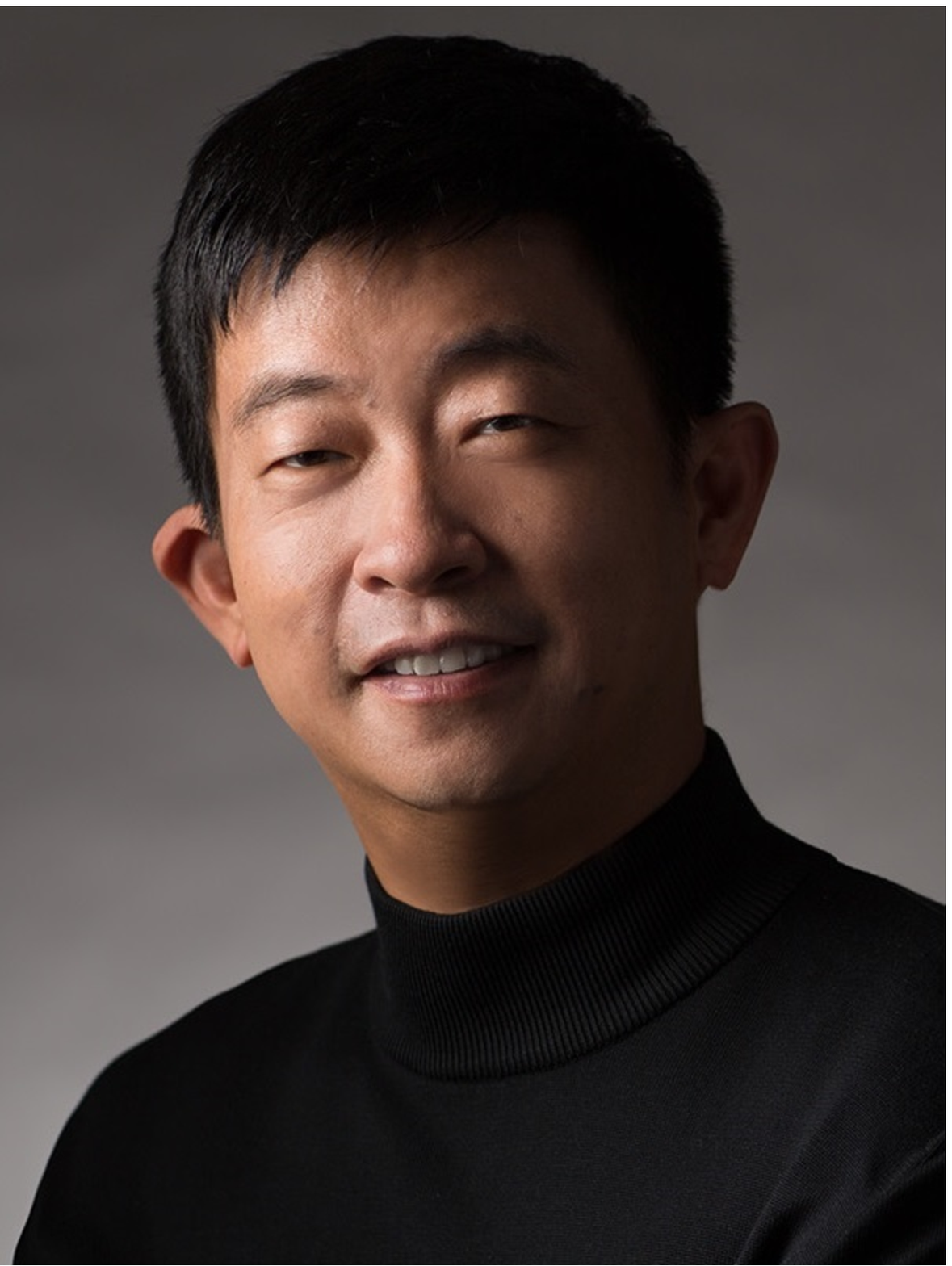}}]{Zhu Han} (S’01–M’04-SM’09-F’14) received the B.S. degree in electronic engineering from Tsinghua University, in 1997, and the M.S. and Ph.D. degrees in electrical and computer engineering from the University of Maryland, College Park, in 1999 and 2003, respectively. 
From 2000 to 2002, he was an R\&D Engineer of JDSU, Germantown, Maryland. From 2003 to 2006, he was a Research Associate at the University of Maryland. From 2006 to 2008, he was an assistant professor at Boise State University, Idaho. Currently, he is a John and Rebecca Moores Professor in the Electrical and Computer Engineering Department as well as in the Computer Science Department at the University of Houston, Texas. Dr. Han’s main research targets on the novel game-theory related concepts critical to enabling efficient and distributive use of wireless networks with limited resources. His other research interests include wireless resource allocation and management, wireless communications and networking, quantum computing, data science, smart grid, security and privacy.  Dr. Han received an NSF Career Award in 2010, the Fred W. Ellersick Prize of the IEEE Communication Society in 2011, the EURASIP Best Paper Award for the Journal on Advances in Signal Processing in 2015, IEEE Leonard G. Abraham Prize in the field of Communications Systems (best paper award in IEEE JSAC) in 2016, and several best paper awards in IEEE conferences. Dr. Han was an IEEE Communications Society Distinguished Lecturer from 2015-2018, AAAS fellow since 2019, and ACM distinguished Member since 2019. Dr. Han is a 1\% highly cited researcher since 2017 according to Web of Science. Dr. Han is also the winner of the 2021 IEEE Kiyo Tomiyasu Award, for outstanding early to mid-career contributions to technologies holding the promise of innovative applications, with the following citation: ``for contributions to game theory and distributed management of autonomous communication networks."
\end{IEEEbiography}
\end{document}